\documentclass[final,authoryear,3p,times,10pt]{elsarticle}
\usepackage{graphicx}
\usepackage{amsmath}
\usepackage{graphicx}
\usepackage{amssymb}
\usepackage{graphicx}
\usepackage{dcolumn}
\usepackage{bm}
\usepackage{color}
\usepackage{graphicx}
\usepackage{multirow}
\usepackage{tabularx}
\usepackage{threeparttable}
\usepackage{booktabs}
\usepackage{multicol}
\usepackage{longtable}
\usepackage{lipsum}
\usepackage{supertabular}
\usepackage{autobreak}

\begin{document}

\begin{frontmatter}

\title{Predicting tail events in a RIA-EVT-Copula framework} 

\author[DF]{Wei-Zhen Li  \fnref{cofirstauthor}}
\author[DF]{Jin-Rui Zhai \fnref{cofirstauthor}} 
\fntext[cofirstauthor]{These authors contribute equally.}
\author[DF,RCE]{Zhi-Qiang Jiang \corref{cor}} \ead{zqjiang@ecust.edu.cn} %
\author[HNU]{Gang-Jin Wang \corref{cor}} \ead{wanggangjin@hnu.edu.cn}
\author[DF,RCE]{Wei-Xing Zhou \corref{cor}} \ead{wxzhou@ecust.edu.cn} %
\cortext[cor]{Corresponding authors.}

\address[DF]{School of Business, East China University of Science and Technology, Shanghai 200237, China}
\address[RCE]{Research Center for Econophysics, East China University of Science and Technology, Shanghai 200237, China}
\address[HNU]{Business School and Center of Finance and Investment Management, Hunan University, Changsha 410082, China} %

\begin{abstract}
Predicting the occurrence of tail events is of great importance in financial risk management. By employing the method of peak-over-threshold (POT) to identify the financial extremes, we perform a recurrence interval analysis (RIA) on these extremes. We find that the waiting time between consecutive extremes (recurrence interval) follow a $q$-exponential distribution and the sizes of extremes above the thresholds (exceeding size) conform to a generalized Pareto distribution. We also find that there is a significant correlation between recurrence intervals and exceeding sizes. We thus model the joint distribution of recurrence intervals and exceeding sizes through connecting the two corresponding marginal distributions with the Frank and AMH copula functions, and apply this joint distribution to estimate the hazard probability to observe another extreme in $\Delta t$ time since the last extreme happened $t$ time ago. Furthermore, an extreme predicting model based on RIA-EVT-Copula is proposed by applying a decision-making algorithm on the hazard probability. Both in-sample and out-of-sample tests reveal that this new extreme forecasting framework has better performance in prediction comparing with the forecasting model based on the hazard probability only estimated from the distribution of recurrence intervals. Our results not only shed a new light on understanding the occurring pattern of extremes in financial markets, but also improve the accuracy to predict financial extremes for risk management. 
\end{abstract}


\end{frontmatter}


\section{Introduction}
\noindent

Recent financial crises like the subprime crisis and the European sovereign debt crisis have motivated the researchers to put considerable efforts on uncovering the risk transmission within different financial institutions or assets \citep{McKibbin-Martin-Tang-2014-JBF, Elyasiani-Kalotychou-Staikouras-Zhao-2015-JFSR, Wang-Xie-Zhao-Jiang-2018-JIFMIM}, different economic sectors \citep{Fonseca-Ignatieva-2018-AE, Shahzad-Bouri-Arreola-Roubaud-Bekiros-2019-AE, Wu-Zhang-Zhang-2019-ES}, different financial markets \citep{Jung-Maderitsch-2014-JBF, Wang-Wu-2019-EE}, and different countries or regions \citep{Kenourgios-Samitas-Paltalidis-2011-JIFMIM, Corradi-Distaso-Fernandes-2012-JEm, Cotter-Suurlaht-2019-EJF} and designing new systematic risk measures to quantitatively evaluate the contribution of individual institution to the systematic risk, such as conditional value at risk (CoVaR) \citep{Adrian-Brunnermeier-2011-AER}, marginal expected shortfall (MES) \citep{Acharya-Pedersen-Philippon-Richardson-2017-RFS}, component expected shortfall (CES) \citep{Banulescu-Dumitrescu-2015-JBF}, SRISK \citep{Acharya-Pedersen-Philippon-Richardson-2017-RFS, Engle-Ruan-2019-PNAS}, and to list a few. However, the above analyses are at least performed on a pair of series. The precondition of applying the proposed methods for predicting risk channels and estimating the systematic risk measures is to precisely forecast the occurrence of the risk events in one asset, one sector, one market, or one country.

In this paper, for a financial series we regard a given value that exceeds a predefined threshold as an extreme risk event, such that for the positive extreme the value should be greater than the threshold and for the negative extreme the value should be smaller than the threshold. Thus, we can uncover the occurring pattern of extremes by investigating the distribution of recurrence intervals corresponding to the waiting time between consecutive extremes. We can also fit the distribution of exceeding size, which is a part above or below the threshold for positive or negative values, by means of the generalized Pareto distribution. Due to the clustering of extremes in financial series \citep{Gresnigt-Kole-Franses-2015-JBF, Jiang-Canabarro-Podobnik-Stanley-Zhou-2016-QF, Jiang-Wang-Canabarro-Podobnik-Xie-Stanley-Zhou-2018-QF}, we may expect that the waiting time between large extremes is relatively small, which indicates the negative correlation between recurrence intervals and exceeding sizes. Our goal is to build a model framework, which is able to incorporate the three stylized facts mentioned above, namely the occurring pattern of extremes, the exceeding size of extremes, as well as their negative correlations. Under this modeling framework, we are able to improve the accuracy of forecasting the future extremes, comparing with the models solely based on the distribution of recurrence intervals \citep{Sornette-Knopoff-1997-BSSA, Bogachev-Eichner-Bunde-2007-PRL, Bogachev-Bunde-2009-PRE, Ludescher-Tsallis-Bunde-2011-EPL, Jiang-Canabarro-Podobnik-Stanley-Zhou-2016-QF, Jiang-Wang-Canabarro-Podobnik-Xie-Stanley-Zhou-2018-QF}. 

The contributions of this paper are listed in the following. First, differing from the strand of literature which separately focuses the distribution of recurrence intervals \citep{Yamasaki-Muchnik-Havlin-Bunde-Stanley-2005-PNAS, Bogachev-Eichner-Bunde-2007-PRL, Bogachev-Bunde-2009-PRE, Li-Wang-Havlin-Stanley-2011-PRE, Xie-Jiang-Zhou-2014-EM, Jiang-Canabarro-Podobnik-Stanley-Zhou-2016-QF} and the distribution of extreme values \citep{Malevergne-Pisarenko-Sornette-2006-AFE, Cumperayot-Kouwenberg-2013-JIMF}, we empirically investigate the joint distribution of recurrence intervals and exceeding sizes and confirm the existence of negative correlations between them. Second, we initially model the joint distribution of recurrence intervals and exceeding sizes by connecting the marginal distributions of recurrence intervals and exceeding sizes with the Archimedean copula functions, including the Frank and Ali-Mikhail-Haq (AMH) copula, which complements the existing studies on recurrence intervals and copulas. The empirical analysis suggests that our model fits the data from the Chinese stock market, the US stock market, and the oil market well. Third, by utilizing the joint distribution model, we are able to derive the hazard probability to observe the extreme in future conditioned on the latest extreme. Therefore, we propose a new model, which considers the characteristics of occurring pattern, exceeding size, and inherent negative correlations between them together to predict the future extremes. Our model overcomes the drawback of the extreme predicting model \citep{Sornette-Knopoff-1997-BSSA, Bogachev-Eichner-Bunde-2007-PRL, Jiang-Canabarro-Podobnik-Stanley-Zhou-2016-QF, Jiang-Wang-Canabarro-Podobnik-Xie-Stanley-Zhou-2018-QF} based on the recurrence interval distribution, which considers the occurring pattern of extremes only.

This paper is organized as follows. Section 2 presents the literature review. Section 3 introduces the data set, and the method to compute the recurrence intervals and exceeding size, as well as the correlation analysis between them. Sections 4 describe the details of the RIA-EVT-Copula model. The application of RIA-EVT-Copula model to forecast extremes is presented in Sections 5. Section 6 concludes.

\section{Literature review}

\subsection{Recurrence interval analysis (RIA) and its application in financial risk management}

Recurrence interval is a type of inter-event time, which measures the waiting time between consecutive extremes. The related studies on recurrence intervals mainly focus on finding suitable distribution formulas for recurrence intervals, bridging the observed distribution form of recurrence intervals with the memory characteristics in original series, and checking the existence of scaling behaviors in recurrence interval distributions. Despite plenty of studies are related to investigating the distribution of recurrence intervals, there are still debates on which distribution can be used to fit the recurrence intervals. On one hand, the power-law distribution is found to well fit the recurrence intervals of the volatilities \citep{Yamasaki-Muchnik-Havlin-Bunde-Stanley-2005-PNAS, Lee-Lee-Rikvold-2006-JKPS, Greco-SorrisoValvo-Carbone-Cidone-2008-PA}, of the returns \citep{Bogachev-Eichner-Bunde-2007-PRL, Bogachev-Bunde-2009-PRE, Ren-Zhou-2010-NJP}, and of the trading volumes \citep{Li-Wang-Havlin-Stanley-2011-PRE, Ren-Zhou-2010-PRE}. On the other hand, the recurrence intervals of the volatilities and of the returns are also reported to conform to a stretched exponential distribution \citep{Wang-Wang-2012-CIE, Xie-Jiang-Zhou-2014-EM, Jiang-Canabarro-Podobnik-Stanley-Zhou-2016-QF, Suo-Wang-Li-2015-EM} and a $q$-exponential distribution \citep{Ludescher-Tsallis-Bunde-2011-EPL, Ludescher-Bunde-2014-PRE, Jiang-Wang-Canabarro-Podobnik-Xie-Stanley-Zhou-2018-QF}, respectively. However, \cite{Jiang-Wang-Canabarro-Podobnik-Xie-Stanley-Zhou-2018-QF} argue that the distribution formula has no influence on the outcomes when applying the recurrence interval analysis to predict financial extremes.  

The apparent distribution form of recurrence intervals is determined by the memory behavior in original series \citep{Chicheportiche-Chakraborti-2013-XXX,Chicheportiche-Chakraborti-2014-PRE}. The memoryless series will generate recurrence intervals following an exponential distribution. Once the series exhibit a characteristic of long linear memory, the obtained recurrence interval distribution will become stretched exponential \citep{Santhanam-Kantz-2008-PRE}. If the series is multifractal, the corresponding recurrence intervals will follow a power-law distribution \citep{Bogachev-Eichner-Bunde-2007-PRL}. Moreover, the distributions of recurrence intervals are not exactly the same for different extreme thresholds, negating the existence of scaling behaviors in recurrence interval distributions. It is found that the recurrence intervals can be fitted by the same distribution but the fitting parameter strongly depends on the extreme threshold \citep{Xie-Jiang-Zhou-2014-EM, Chicheportiche-Chakraborti-2014-PRE, Suo-Wang-Li-2015-EM, Jiang-Canabarro-Podobnik-Stanley-Zhou-2016-QF}. More surprisingly, the recurrence interval distribution only depends on the threshold, and not on a specific asset or on a time resolution \citep{Ludescher-Tsallis-Bunde-2011-EPL, Ludescher-Bunde-2014-PRE,  Jiang-Canabarro-Podobnik-Stanley-Zhou-2016-QF}.

Very limit research is concentrated on applying the recurrence interval analysis on financial risk management, like estimation of risk measures and prediction of extremes. By linking thresholds with quantiles,  \cite{Bogachev-Bunde-2009-PRE, Ludescher-Tsallis-Bunde-2011-EPL} propose a VaR estimating method based on the recurrence interval analysis and advocate that their new method is superior to the nonparametric methods based on the overall or local return distributions. By defining a conditional hazard probability based on the recurrence interval distribution, one is able to probability predict the occurrence of future extremes \citep{Sornette-Knopoff-1997-BSSA, Bogachev-Eichner-Bunde-2007-PRL, Ren-Zhou-2010-NJP, Xie-Jiang-Zhou-2014-EM}. Such an extreme predicting method outperforms the precursory pattern recognition technique when training series are multifractal \citep{Bogachev-Bunde-2009-PRE}. However, the conclusions in the above mentioned studies are drawn from in-sample analysis. Recently, \cite{Jiang-Canabarro-Podobnik-Stanley-Zhou-2016-QF, Jiang-Wang-Canabarro-Podobnik-Xie-Stanley-Zhou-2018-QF} find that extreme forecasting methods based on the recurrence interval analysis does have predicting power in out-of-sample tests, but not as good as the results of in-sample tests. 

\subsection{Copula and its application in financial risk management}

According to the theorem proposed by \cite{Sklar-1959-PISUP}, a multivariate distribution function can be decomposed into its univariate marginal distributions and a copula function which completely captures the dependence within random variables. Obviously, such decomposition can simplify the modeling of multivariate distributions, which enables us to model the marginal distributions and multivariate dependence structure separately. The advantages of using copula functions are that no additional restriction is needed for the marginal distributions and many kinds of dependence structures, including linearity, non-linearity, upper tail, lower tail, asymmetric tail, and so on, can be incorporated.  The copula has important applications in finance due to its easy implementation, such as modelling the dependence between financial series, improving the accuracy of estimating risk measures and further to better portfolio allocations, uncovering the risk spillover between different markets, and to list a few.  

According to the definition of copula, a direct application of copula is to measure and model the dependence structure between financial series. By fitting the data with different copula functions, we are able to find the best copulas to describe the dependence structure and the copula parameter offers a dependence measure. Comparing with the conventional Pearson correlation, copula has the advantages of capturing the tail, asymmetric, and time-varying dependence. In the empirical analysis, the copulas are widely applied to uncover the cross-market dependence and give incompatible results. For example, between stock and exchange markets, \cite{Ning-2010-JIMF} finds that both upper and lower tail dependence are symmetric and \cite{Wang-Wu-Lai-2013-JBF} reports that only upper tail dependence is symmetric. The copula model with its fitted parameters also offer an avenue to model to the tail, asymmetric, and time-varying dependence. \cite{Nikoloulopoulos-Joe-Li-2012-CSDA} employ the vine copulas with flexible tail dependence to model the tail asymmetries in financial returns. By extending the unconditional copula to the conditional case, \cite{Patton-2006-IER} successfully models the time-varying asymmetric dependence structure between exchange returns. Furthermore, copula also provides an efficient way to model the dependence in high dimensions. \cite{Oh-Patton-2017-JBES} propose a factor copula model which has the capacity to give reliable results from the simulating and empirical tests involving up to 100 variables.

Better dependence modeling with copulas, of course, gives rise to higher accurate estimation of portfolio risk measures. \cite{Siburg-Stoimenov-Weiss-2015-JBF} report that the portfolio VaR forecasting model based on the copulas with lower tail dependence generates fewer VaR exceedances in out-of-sample tests. More importantly, a better estimation of risk measures can further improve portfolio optimizations. Specifically, the linear correlation measure (correlation coefficient) and risk measure (variance) in the classical mean risk framework can be replaced with the nonlinear correlation measure (copula parameters) and tail risk measure based on copulas. \cite{Boubaker-Sghaier-2013-JBF} demonstrate that this updated approach based on copulas outperforms the classical mean-variance approach. \cite{Chu-2011-JBF} also finds that asset allocation based on the copula dependence results in statistically and economically significant gains. However, the portfolio diversification based on copulas is not always efficient. \cite{Chollete-delaPena-Lu-2011-JBF} explore the profits of optimized diversification based on different dependence measures and find that there is no significant difference in portfolio gains between copula and correlation, which means that copula also has shortcomings to describe some complex dependence structures in international stock markets.  

The copula has also been applied to investigate the risk spillover effects due to its advantage of dependence modeling. This strand of studies are mainly based on the systematic risk measure CoVaR, defined as the risk measure (VaR) of a financial system conditioned on being in financial distress \citep{Adrian-Brunnermeier-2011-AER} and further is extended to measure the contribution to systematic risk for one institution in financial distress \citep{Girardi-Ergun-2013-JBF}. And based on copulas a two-step procedure was proposed to estimate the CoVaR, which enables us to take the tail dependence into account in particular \citep{Reboredo-Ugolini-2015-JIMF}.  The CoVaR-copula method is applied to study the dynamic and asymmetric risk spillover effects between return and implied volatility index in the oil markets \citep{Liu-Ji-Fan-2017-EE} and between oil prices and exchange rates \citep{Ji-Liu-Fan-2019-EE}. Such CoVaR-copula approach is further updated by incorporating the behavior of regime shift dependence to reveal the risk spillover between energy and agricultural commodity markets \citep{Ji-Bouri-Roubaud-Shahzad-2018-EE} and by combining with the variational mode decomposition method to detect the risk dependence structure between oil markets and stock markets \citep{Mensi-Hammoudeh-Shahzad-Shahbaz-2017-JBF, Li-Wei-2018-EE}.

\section{Data sets}

\begin{figure}[htbp]
\centering
\includegraphics[width=10cm]{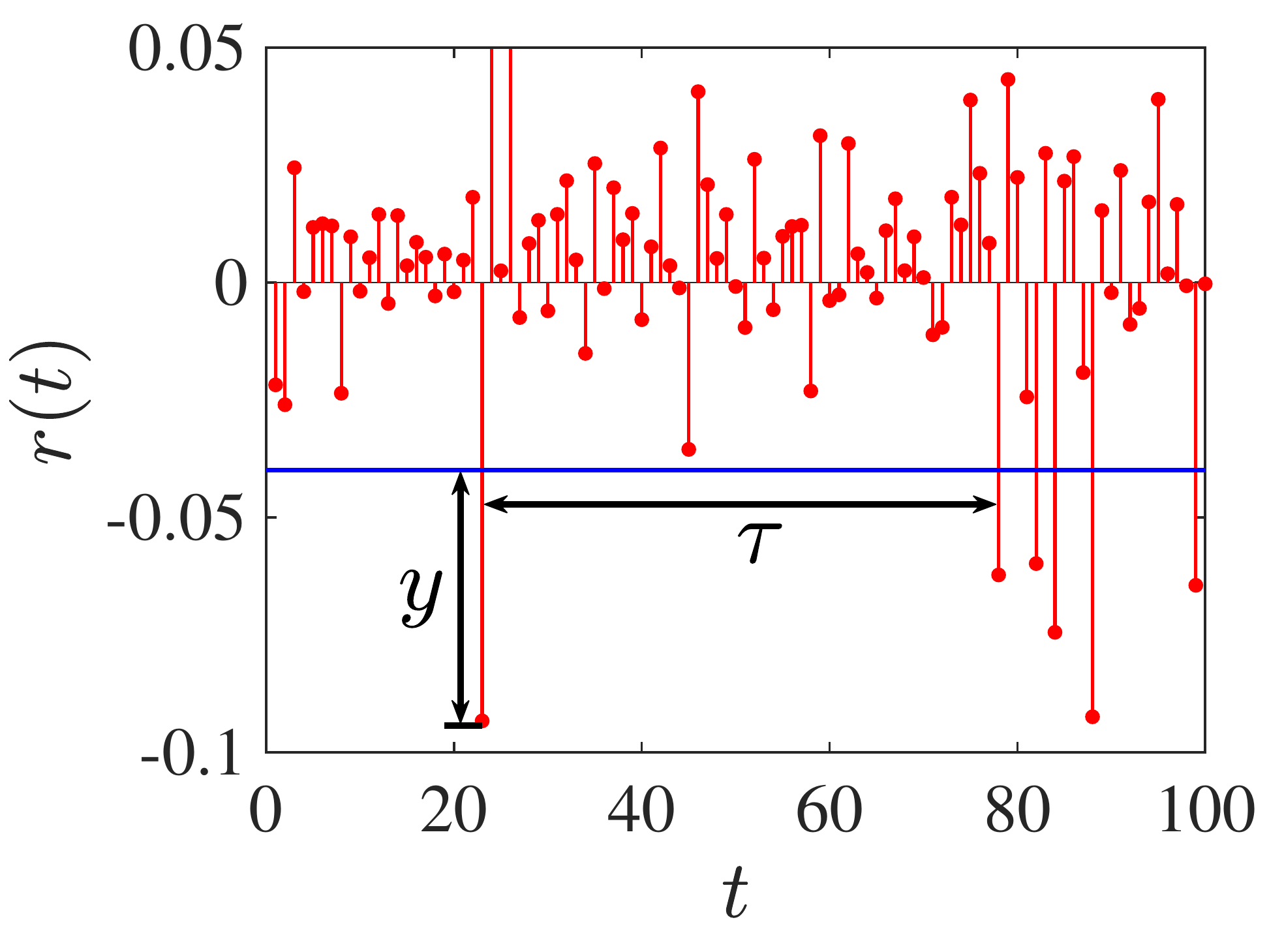}
\caption{\label{Fig:DFN:Tau:EY} Definition of the recurrence interval $\tau$ and exceeding size $y$. }
\end{figure}

Our analysis is based on three daily data sets, including the Shanghai Stock Exchange Composite (SSEC) index spanning from January 1, 1997 to December 31, 2019, the WTI Crude Oil Prices (WTI) price recorded March 31, 1983 to December 31, 2019, and the Dow Jones Industrial Average (DJIA) index covering from February 16, 1885 to December 31, 2019. The daily return $r(t)$ is defined as, 
\begin{equation}
  r(t) = \ln I(t) - \ln I(t-1).
  \label{Eq:Return}
\end{equation}
where $I(t)$ is the index on day $t$. We employ the method of peak over threshold (POT) to identify extremes in returns, which are quantitatively described by its exceeding size and recurrence interval. As shown in Fig.~\ref{Fig:DFN:Tau:EY}, the recurrence interval $\tau$ captures the waiting time between consecutive extremes and the exceeding size $y$ measures the value above and below the threshold for positive and negative returns. 

Table~\ref{Tab:Stat:Tau:EY} lists the descriptive statistics of the recurrence interval $\tau$ and exceeding size $y$ associated with the threshold of 90\% (respectively, 10\%) quantile for the positive (negative) returns of SSEC, DJIA, and WTI. One can see that for the recurrence intervals and exceeding sizes of both thresholds the mean is significantly greater than the median for the three indexes, indicating the mean is strongly influenced by the extreme values. We find that the average recurrence interval $\langle \tau \rangle$ and the average exceeding size $\langle y \rangle$ satisfy $\langle \tau \rangle_{\rm{DJIA}} > \langle \tau \rangle_{\rm{SSEC}} > \langle \tau \rangle_{\rm{WTI}}$ and $\langle y \rangle_{\rm{DJIA}} < \langle y\rangle_{\rm{SSEC}} < \langle y \rangle_{\rm{WTI}}$, indicating that the volatile degree of the three markets follows WTI $>$ SSEC $>$ DJIA. We can also find that the recurrence intervals and exceeding sizes are right-skewed and fat-tailed, supported by that their skewness is positive and kurtosis is greater than 3. This is in consistent with that the recurrence intervals of extreme returns conform to the $q$-exponential distribution or the stretched exponential distribution \citep{Jiang-Canabarro-Podobnik-Stanley-Zhou-2016-QF, Jiang-Wang-Canabarro-Podobnik-Xie-Stanley-Zhou-2018-QF}, and the extreme values of financial returns follow the generalized pareto distribution (GPD) \citep{Cumperayot-Kouwenberg-2013-JIMF}.

\begin{table*}[htbp]
\setlength\tabcolsep{3.3pt}
\footnotesize
\centering
 \caption{\label{Tab:Stat:Tau:EY} Descriptive statistics of the recurrence interval $\tau$ and exceeding size $y$. This table reports number of observations (obsv), mean, median, standard deviation (stdev), skewness (skew), and kurtosis (kurt) of the recurrence intervals and exceeding sizes of extreme events. The extreme events correspond to the financial returns exceeds the quantile threshold of 10\% and 90\% for negative and positive returns, respectively. Panels A–C present the results from the data sets of SSEC, DJIA, and WTI.}
 \medskip
\begin{tabular}{rrrrrrcrrrrrrrrrrrr}
\toprule 
& &   \multicolumn{ 8}{c}{Recurrence interval $\tau$} &&   \multicolumn{8}{c}{Exceeding size $y$}  \\
\cline{3-10}\cline{12-19}$Q$ & & obsv & mean & max & min & median & stdev & skew & kurt & & obsv & mean & max &  min & median & stdev & skew & kurt  \\
\midrule
\multicolumn{ 19}{l}{Panel A: Fits to SSEC from January 1, 1997 to December 31, 2019} \\
90\% & & 434 & 9.606 & 187 & 1 & 5 & 15.134 & 5.960 & 55.742 & & 434 & 0.011 & 0.076 & 2.36E-05 & 0.008 & 0.012 & 2.535 & 10.793\\
10\% & & 458 & 9.063 & 74 & 1 & 4 & 10.947 & 2.210 & 8.482 & & 458 & 0.012 & 0.077 & 6.60E-06 & 0.007 & 0.014 & 2.035 & 7.620\\
\multicolumn{ 19}{l}{Panel B: Fits to DJIA from February 16, 1885 to December 31, 2019} \\
90\% & & 3589 & 9.997 & 394 & 1 & 5 & 16.546 & 7.444 & 111.329 & & 3589 & 0.007 & 0.132 & 2.66E-07 & 0.004 & 0.010 & 4.220 & 32.221\\ 
10\% & & 3587 & 9.989 & 302 & 1 & 4 & 16.288 & 5.076 & 50.699 & & 3587 & 0.009 & 0.246 & 3.29E-07 & 0.005 & 0.012 & 5.127 & 65.381\\ \midrule\multicolumn{ 19}{l}{Panel C: Fits to WTI from March 31, 1983 to December 31, 2019} \\
90\% & & 888 & 10.070 & 264 & 1 & 5 & 18.329 & 6.953 & 72.857 & & 888 & 0.016 & 0.139 & 1.61E-05 & 0.010 & 0.020 & 2.542 & 11.001\\ 
10\% & & 961 & 9.486 & 230 & 1 & 5 & 16.106 & 6.851 & 70.938 & & 961 & 0.018 & 0.376 & 5.46E-07 & 0.011 & 0.023 & 5.468 & 63.834\\  \midrule                                                                                                                                                                                                    
\bottomrule
\end{tabular} 
\end{table*}

\begin{figure}[h]
 \centering
\includegraphics[width=16cm]{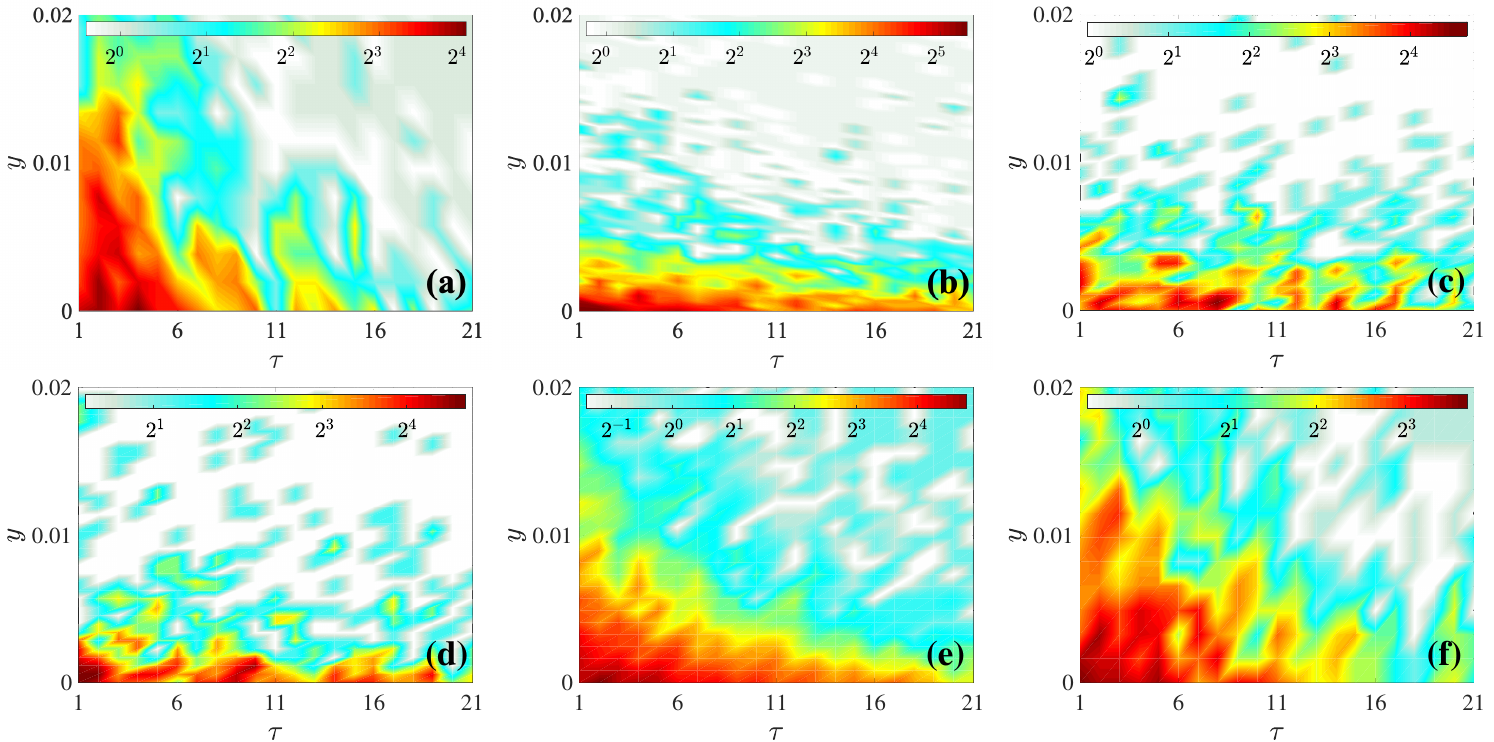}     
 \caption{\label{Fig:BEPDF:Tau:EY} Empirical joint probability distribution of the recurrence interval $\tau$ and exceeding size $y$ for SSEC (a), DJIA (b), WTI (c) with the quartile of 90\% and for SSEC (c), DJIA(d), WTI(e) with the quartile of 10\%.}
\end{figure}

According to the stylized fact that financial extremes are clustered, one can infer that the recurrence intervals and exceeding sizes are negatively correlated. Fig.~\ref{Fig:BEPDF:Tau:EY} illustrates the empirical joint probability distribution of recurrence intervals $\tau$ and exceeding sizes $y$ for the three indexes with the quantile threshold of 90\% and 10\%. We find that the probability to observe the large exceeding size decreases with the increment of the recurrence intervals, which supports the negative correlation between $\tau$ and $y$. Table~\ref{Tab:Cor:Tao:EY} reports the correlation coefficient $\rho$ and its $p$-value between recurrence interval $\tau$ and exceeding size $y$ for the three indexes with different thresholds. It is observed that 5/6 of the correlation coefficients are negative and statistically significant at the level of 0.05. However, such an important negative correlation is not considered in the RI-based risk models \citep{Jiang-Wang-Canabarro-Podobnik-Xie-Stanley-Zhou-2018-QF}.

\begin{table}[htbp]
\setlength\tabcolsep{6pt}
\footnotesize
\centering
 \caption{\label{Tab:Cor:Tao:EY} Correlation coefficient $\rho$ and its $p$-value between recurrence intervals $\tau$ and exceeding sizes $y$ of the extremes with different thresholds for SSEC, DJIA, and WTI.}
 \medskip
  \begin{threeparttable}
 \begin{tabular}{rcllcllcll}
\toprule
& & \multicolumn{ 2}{c}{SSEC} & & \multicolumn{ 2}{c}{DJIA} & & \multicolumn{ 2}{c}{WTI} \\
\cline{3-4}\cline{6-7}\cline{9-10}
& & \multicolumn{1}{c}{$\rho$} & \multicolumn{1}{c}{$p$-value} &  &  \multicolumn{1}{c}{$\rho$} & \multicolumn{1}{c}{$p$-value} & & \multicolumn{1}{c}{$\rho$} & \multicolumn{1}{c}{$p$-value} \\
\midrule
90\% & & -0.053 & 0.200 & & -0.102$^{***}$ & 0.000 & & -0.080$^{**}$ & 0.034  \\
10\% & & -0.122$^{***}$ & 0.002 & & -0.171$^{***}$ & 0.000 & & -0.097$^{***}$ & 0.009  \\
\bottomrule
\end{tabular} 
\begin{tablenotes} 
\footnotesize
\item ${*}$, ${ **}$, and ${***}$ represent the significant level of 10\%, 5\%, and 1\%, respectively.
\end{tablenotes}
\end{threeparttable}
\end{table}

\section{RIA-EVT-Copula model}

\subsection{Model construction}

We employ the copula function to model the joint distribution of recurrence interval and exceeding size, which allows us to not only take into account the fat-tailed nature of marginal distributions, but also consider the potential tail dependence between recurrence interval and exceeding size. The joint density $f(\tau, y)$ is given as follows:
\begin{equation}
  f(\tau, y) = p(\tau) g(y) c(u, v),
  \label{Eq:JointDis:Tau:Ey}
\end{equation}
where $p(\tau)$ and $g(y)$ are the probability densities of recurrence intervals and exceeding sizes. $c(u, v)$ is the copula density function, in which $u = P(\tau) = \int_{-\infty} ^{\tau} p(x) {\rm{d}} x $ and $v = G(y) = \int_{-\infty} ^{y} g(x) {\rm{d}} x $.
 
For the recurrence intervals, $p(\tau)$ can be modeled by the following distributions, including the stretched exponential distribution \citep{Wang-Wang-2012-CIE,Suo-Wang-Li-2015-EM, Jiang-Canabarro-Podobnik-Stanley-Zhou-2016-QF},
\begin{equation}
  p_{\rm{SE}}(\tau)=a e^{[-(b\tau)^{\mu}]},
  \label{Eq:Tau:StchExp}
\end{equation}
the $q$-exponential distribution \citep{Ludescher-Bunde-2014-PRE, Ludescher-Tsallis-Bunde-2011-EPL, Chicheportiche-Chakraborti-2014-PRE}, 
\begin{equation}
  p_{\rm{qE}}(\tau)=(2-q)\lambda[1+(q-1)\lambda\tau]^{-\frac{1}{q-1}},
  \label{Eq:Tau:qExp}
\end{equation}
and the Weibull distribution \citep{Reboredo-Rivera-Castro-Machado-2014-QF},
\begin{equation}
  p_{\rm{W}}(\tau)=\frac{\alpha}{\beta} \left(\frac{\tau}{\beta}\right)^{\alpha-1} e^{\left[-\left(\frac{\tau}{\beta}\right)^{\alpha} \right]}.
  \label{Eq:Tau:Weibull}
\end{equation}

Following \cite{Cumperayot-Kouwenberg-2013-JIMF}, the generalized Pareto distribution (GPD) is used to model the distribution of exceeding size $g(y)$, which can be written as, 
\begin{equation}
\begin{cases} 
g(y)=\frac{1}{\varphi}(1+\xi\frac{y}{\varphi})^{-\frac{1}{\xi}-1} , ~\xi\neq0, \\
g(y)=\frac{1}{\varphi}\exp(-\frac{y}{\varphi}) , ~~~~~~~~\xi=0.
\end{cases}
\label{Eq:Ey:GPD}
\end{equation}

There are many types of copula functions, such as binary normal copula function, t-copula function, and Archimedean copula functions. In this paper, we employ the Archimedean copula function to connect the distributions of recurrence intervals and exceeding sizes. The widely used Archimedean functions are the Clayton, Gumbel, Frank, and Ali-Mikhail-Haq (AMH) copulas. Since Clayton and Gumbel copula only capture the asymmetric dependence, we thus employ Frank copula and AMH copula to model $c(u, v)$, which are given in the following, 
\begin{equation}
  C_{\rm{Frank}}(u,v;\theta_f)=-\frac{1}{\theta_f}\ln \left[ 1+\frac{(e^{-{\theta_f}u}-1)(e^{-{\theta_f}v}-1)}{e^{-\theta_f}-1} \right],
  \label{Eq:copula:Frank}
\end{equation}
\begin{equation}
  C_{\rm{AMH}}(u,v;\theta_a)=\frac{uv}{1-\theta_a(1-u)(1-v)}.
  \label{Eq:copula:AMH}
\end{equation}
And their density functions are
\begin{equation}
  c_{\rm{Frank}}(u,v;\theta_f) = \frac{\theta_f(1-e^{-\theta_f})e^{-\theta_f(u+v)}}{[(1-e^{-\theta_f})-(e^{{-\theta_f}u}-1)(e^{-{\theta_f}v}-1)]^2},
  \label{Eq:copula:Frank:density}
\end{equation}
\begin{equation}
  c_{\rm{AMH}}(u, v; \theta_a)=\frac{1+\theta_a[(1+u)(1+v)-3]+{\theta_a}^{2}(1-u)(1-v)}{[1-\theta_a(1-u)(1-v)]^3} ,
  \label{Eq:coppula:AMH:density}
\end{equation}
where $\theta_f$ and $\theta_a$ are the Frank and AMH copula parameters, which satisfy the following inequations: $\theta_f \neq 0$ and $-1 \le \theta_a < 1$. 

\subsection{Parameter estimation}

The parameters of RIA-EVT-Copula model are estimated via the method of inference functions for margins. We first separately estimate the parameters of recurrence interval distribution and the exceeding size distribution by the maximum likelihood estimation (MLE) method. Once we have the distribution parameters, we then estimate the copula parameters by maximizing the likelihood function of copula functions.

Following \cite{Jiang-Wang-Canabarro-Podobnik-Xie-Stanley-Zhou-2018-QF}, for recurrence intervals, the parameters of the stretched exponential distribution, $q$-exponential distribution, and Weibull distribution are determined through maximizing the following log-likelihood functions, such that 
\begin{equation}
  {\ln}L_{\rm{SE}}=n{\ln}a-\sum_{i=1}^{n}(b\tau_{i})^{\mu},~~a=\frac{\mu\Gamma(\frac{2}{\mu})}{\Gamma(\frac{1}{\mu})^{2}\tau_Q}, ~~b=\frac{\Gamma(\frac{2}{\mu})}{\Gamma(\frac{1}{\mu})\tau_Q}, 
  \label{Eq:lnL:Tau:StchExp}
\end{equation}
\begin{equation}
  {\ln}L_{\rm{qE}}=n{\ln}[\lambda(2-q)]-\frac{1}{q-1}\sum_{i=1}^{n}{\ln}\left[1+(q-1)\lambda\tau_{i}\right],~~\lambda=\frac{1}{\tau_{Q}(3-2q)},
  \label{Eq:lnL:Tau:qExp}
\end{equation}
\begin{equation}
  {\ln}L_{\rm{W}}=n{\ln}\frac{\alpha}{\beta}+\sum_{i=1}^{n}\left[(\alpha-1){\ln}\frac{\tau_{i}}{\beta}-\left(\frac{\tau_{i}}{\beta}\right)^{\alpha}\right],~~\beta=\frac{\tau_{Q}}{\Gamma(1+\frac{1}{\alpha})},
  \label{Eq:LnL:Tau:Weibull}
\end{equation}
where $\tau_Q$ is the average value of the recurrence intervals. One can see that there is only one parameter to estimated in each log-likelihood function. We thus simply discretize the parameters in their value range, that is $\mu \in (0, 1)$ for the stretched exponential distribution, $q \in (0, 1.5)$ for the $q$-exponential distribution, and $\alpha \in (0, 1)$ for the Weibull distribution, with a step of $10^{-6}$ and calculate their corresponding log-likelihood functions. The best estimations of $\mu$, $q$, and $\alpha$ are associated with the maximum value of ${\ln}L_{\rm{SE}}$,  ${\ln}L_{\rm{qE}}$, and ${\ln}L_{\rm{W}}$.

For the distribution of exceedances, there are two parameters $\xi$ and $\varphi$ to be estimated in the GPD. By giving the following log-likelihood function, 
\begin{equation}
  \ln L_{\rm{GPD}}=-n\ln \varphi -\left(\frac{1}{\xi }+1\right)\sum_{i=1}^n\ln \left(1+\xi \frac{y_i}{\varphi}\right).
  \label{Eq:lnL:Ey:GPD}
\end{equation}
The best estimation of $\xi$ and $\varphi$ is obtained by maximizing $\ln L_{\rm{GPD}}$. 

The log-likelihood function of the Frank and AMH copula can be written as follows, 
\begin{equation}
  \ln L_{\rm{Frank}}=n\ln \theta_f \left(1-e^{-\theta_f }\right)-{\theta_f}\sum_{i=1}^n (u_i+v_i)-2\sum_{i=1}^n\ln \left[\left(1-e^{-\theta_f} \right)- \left(e^{-{\theta_f}u_i}-1 \right) \left(e^{-{\theta_f}v_i}-1 \right)\right],
  \label{Eq:lnL:copula:Frank}
\end{equation}
\begin{equation}
  \ln L_{\rm{AMH}}=\sum_{i=1}^n \ln \left[1+\theta_a (1+u_i)(1+v_i)-3\theta_a+\theta_a^{2}(1-u_i)(1-v_i)\right]-3 \sum_{i=1}^n\ln \left[1-\theta_a(1-u_i)(1-v_i) \right].
  \label{Eq:lnL:copula:AMH}
\end{equation}
When we have the distribution parameters of the recurrence intervals and exceeding size, we can calculate the $u_i$ and $v_i$ for each $\tau_i$ and $y_i$,  which allows us to further estimate the copula parameters $\theta_f$ and $\theta_a$ by maximizing $\ln L_{\rm{Frank}}$ and $\ln L_{\rm{AMH}}$ for the Frank and AMH copula, respectively. 

Following \cite{Zhang-2005-PhDthesis}, we adopt the root mean square error (RMSE) and the Akaike information criterion (AIC) to indicate the goodness of fits. The RMSE and AIC are defined as follows,
\begin{equation}
  {\rm RMSE}=\sqrt{\frac{1}{n-1}\sum_{i=1}^{n}\left(F^e_{i}-F^t_{i}\right)^{2}}
  \label{Eq:RMSE}
\end{equation}
\begin{equation}
  {\rm AIC}=n\ln \left[ \frac{1}{n-1}\sum_{i=1}^{n}\left(F^e_{i}-F^t_{i}\right)^2 \right]+2m
  \label{Eq:ln_AIC}
\end{equation}
where $n$ is the sample size, $m$ is the number of model parameters, and $F^e_{i}$ and $F^t_{i}$ correspond to the empirical and theoretical joint distribution of the recurrence intervals and exceeding sizes. The smaller RMSE and AIC value are, the better fits the copula function has.

\subsection{Fitting results}

\begin{figure}[htbp]
 \centering
   \includegraphics[width=16cm]{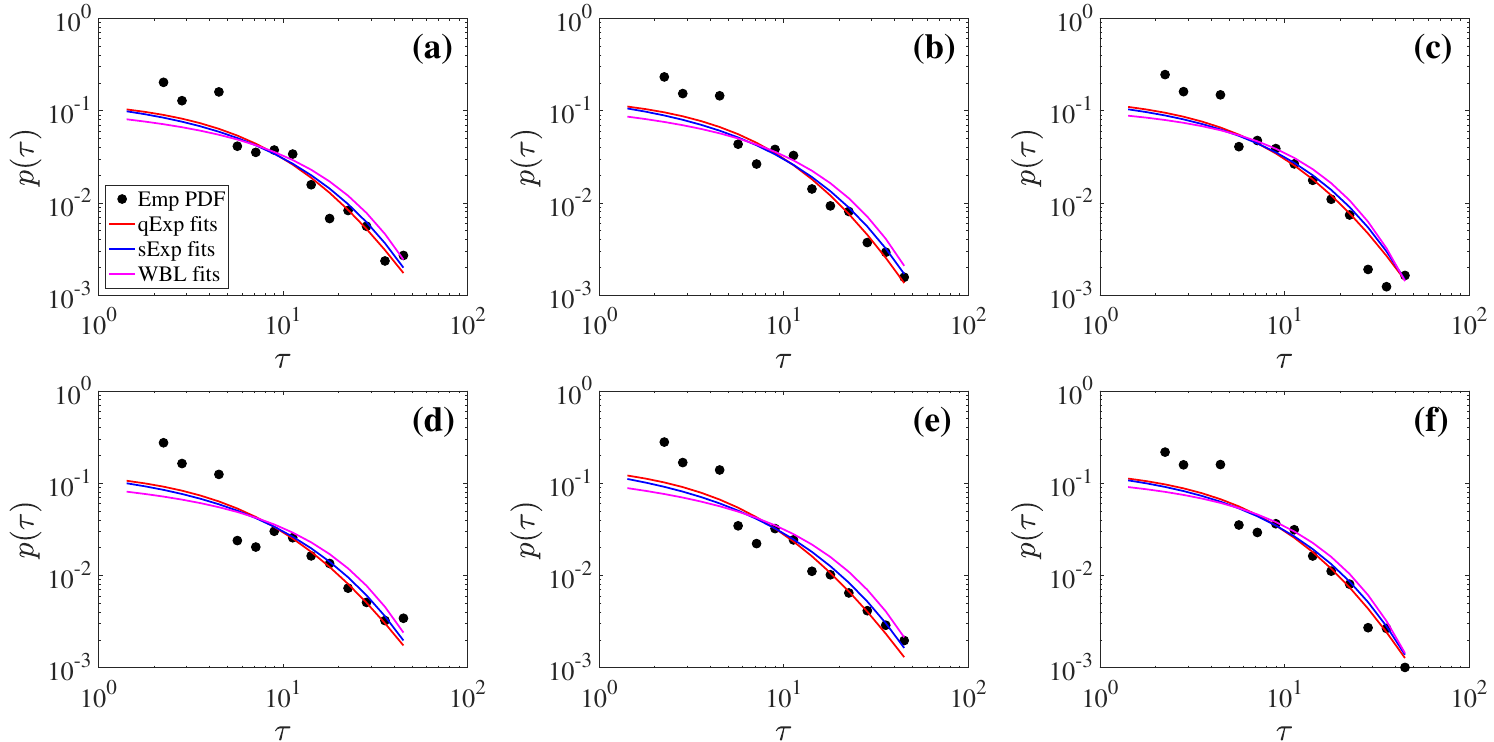}
 \caption{\label{Fig:PDF:Tau:Fits} Probability distribution of the recurrence intervals $\tau$ for the positive extremes of SSEC (a), DJIA (b), WTI (c) above the quantile threshold of 90\% and for the negative extremes of SSEC (d), DJIA (e), WTI (f) below the quantile of 10\%. The filled markers are empirical distributions and the solid lines are the best fits to candidate distributions, including the stretched exponential, the $q$-exponential distribution, and the Weibull distribution. }
\end{figure}

In order to test the power of the above fitting technology, we calibrate the RIA-EVT-Copula model with the three indexes in their whole sample. Fig.~\ref{Fig:PDF:Tau:Fits} illustrates the distributions of recurrence intervals for the positive returns of three indexes with quantile thresholds 90\% and 10\%. One can see that the markers and the solid lines are in good agreement with each other in each panel. In comparison of the three fits, the $q$-exponential distribution gives the best fits. We also list the results, including the distribution parameters and the logarithm likelihood functions, of fitting the recurrence intervals $\tau$ to the three candidate distributions in Table~\ref{Tb:FitModel:Pars}. We find that for each distribution the fitting parameters of different indexes are very close to each other, which are in agreement with the parameter fitting values reported by \cite{Jiang-Canabarro-Podobnik-Stanley-Zhou-2016-QF}. We also observe that the $q$-exponential distribution gives the best fits to the recurrence intervals, evidenced by the observation that the logarithm likelihood function of $q$-exponential distribution is the maximum. 

\begin{figure}[htbp]
 \centering
   \includegraphics[width=16cm]{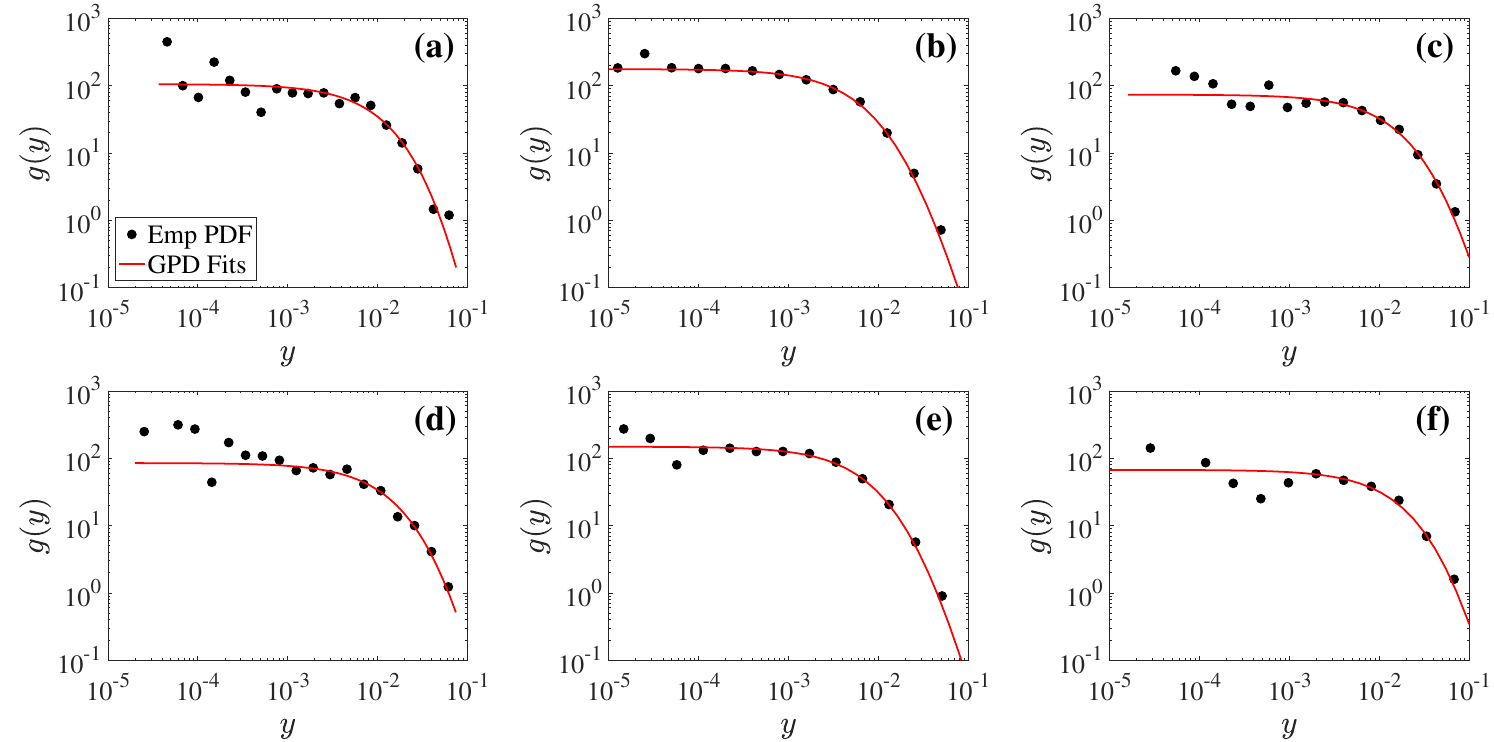}
 \caption{\label{Fig:PDF:y:Fits} Probability distribution of the exceedances $y$ for the positive extremes of SSEC (a), DJIA (b), WTI (c) above the quantile threshold of 90\% and for the negative extremes of SSEC (d), DJIA (e), WTI (f) below the quantile of 10\%. The filled markers are empirical distributions and the solid lines are the best fits to the generalized Pareto distribution.}
\end{figure}

We further compute the empirical distributions of the exceedances $y$ and fit them to the generalized Pareto distribution by MLE. Fig.~\ref{Fig:PDF:y:Fits} illustrates the empirical distribution of the exceedance $y$ and its best fits to the generalized Pareto distribution for the extremes obtained from quantile thresholds of 90\% and 10\%. It is observed that the empirical distributions and the fitting distributions agree with each other very well, especially in the tail part. The fitting distribution parameters are also listed in Table~\ref{Tb:FitModel:Pars}. We also find that there exists to be $\xi_{\rm{DIJA}} > \xi_{\rm{WTI}} > \xi_{\rm{SSEC}}$, indicating that the extremes of SSEC have the smallest value. This may result from the policy of price limits in the Chinese stock markets. 

Once the distribution parameters of the recurrence intervals and the exceedances are obtained, we can thus estimate the parameters of two copula functions. The fitting results are also listed in Table~\ref{Tb:FitModel:Pars}. We find that the estimating parameters of both copula functions are all less than zero, indicating the existence of negative correlations between recurrence intervals and exceeding sizes. For SSEC and WTI, we find that the AMH copula is the best copula function to capture the correlation between $\tau$ and $y$ obtained from positive and negative extremes. For DJIA, we observe that the Frank (respectively, the AMH) copula is the best copula function to describe the dependence between $\tau$ and $y$ obtained from the positive (negative) extremes. 

\begin{table}[htbp]
\setlength\tabcolsep{2pt}
\footnotesize
\centering\tiny\small
\caption{\label{Tb:FitModel:Pars} Fitting results of the RIA-EVT-Copula model for SSEC, DJIA, and WTI. This table reports distribution parameters and likelihood functions  ($\mu$ and $\ln L_{\rm{sE}}$ for the stretched exponential distribution, $q$ and $\ln L_{\rm{qE}}$ for the $q$-exponential distribution, and $\alpha$ and $\ln L_{\rm{W}}$ for the Weibull distribution) for recurrence intervals, distribution parameters of the GPD ($\varphi$ and $\xi$) for exceeding sizes, copula parameters ($\theta_f$ for the Frank copula, and $\theta_a$ for the AMH copula), and good of fitness (RMSE and AIC) for copula functions. For the fits of recurrence interval distributions, the distribution parameter with the maximum likelihood function is highlighted in bold. For the fits of copula functions, the copula parameter with the minimum RMSE and AIC is highlighted in bold.}
\medskip
 \centering
 \begin{tabular}{rcccccccccccccccccccc}
\toprule
&& \multicolumn{8}{c}{Recurrence intervals} && \multicolumn{2}{c}{Exceedances} && \multicolumn{7}{c}{Copula functions}\\
\cline{3-10} \cline{12-13} \cline{15-21}
&& \multicolumn{2}{c}{stret. exp.} && \multicolumn{2}{c}{$q$-exp.} &&\multicolumn{2}{c}{Weibull} && \multicolumn{2}{c}{GPD} && \multicolumn{3}{c}{Frank} && \multicolumn{3}{c}{AMH}\\
\cline{3-4}\cline{6-7} \cline{9-10} \cline{12-13} \cline{15-17} \cline{19-21}
$Q$ && $\mu$ & $\ln L_{\rm{sE}}$ && $q$ & $\ln L_{\rm{qE}}$ && $\alpha$ & $\ln L_{\rm{W}}$ && $\varphi$ & $\xi$ && $\theta_f$ & RMSE & AIC &&  $\theta_a$& RMSE & AIC \\
\midrule
\multicolumn{21}{l}{Panel A: Fits to SSEC from January 1, 1997 to December 31, 2019} \\
90\% && 0.72 & -1402.1 && {\bf1.21} & -1392.3 && 0.93 & -1412.7 && 0.01 &0.11 && -1.545 & 0.027 & -3128.9 && {\bf-0.778} & 0.026 & -3179.4 \\
10\% && 0.75 & -1459.4 && {\bf1.19} & -1457.3 && 0.95 & -1466.0 && 0.01 &0.13 && -1.224 & 0.036 & -3035.6 && {\bf-0.676} & 0.036 & -3049.9 \\
\midrule
\multicolumn{21}{l}{Panel B: Fits to DJIA from February 16, 1885 to December 31, 2019} \\
90\% && 0.64 & -8529.1 && {\bf 1.25} & -8463.7 && 0.89 & -8626.9 && 0.01 &0.26 && {\bf-1.182} & 0.022 & -19941.5 && -0.712 & 0.023 & -19718.5 \\
10\% && 0.59 & -8655.3 && {\bf 1.29} & -8591.1 && 0.86 & -8775.9 && 0.01 &0.24 && -2.061 & 0.027 & -19322.8 && {\bf-1.000} & 0.023 & -20292.9 \\
\midrule
\multicolumn{21}{l}{Panel C: Fits to WTI from March 31, 1983 to December 31, 2019} \\
90\% && 0.72 & -2259.6 && {\bf 1.22} & -2237.9 && 0.95 & -2280.3 && 0.01 &0.18 && -1.175 & 0.031 & -4892.2 && {\bf-0.607} & 0.030 & -4924.3 \\
10\% && 0.70 & -2327.1 && {\bf 1.22} & -2311.1 && 0.93 & -2348.0 && 0.02 &0.20 && -1.507 & 0.029 & -5144.8 && {\bf-0.760} & 0.028 & -5223.2 \\ 
\bottomrule
\end{tabular}  
\end{table}

\section{Forecasting extreme events}

Following \cite{Jiang-Canabarro-Podobnik-Stanley-Zhou-2016-QF, Jiang-Wang-Canabarro-Podobnik-Xie-Stanley-Zhou-2018-QF}, we employ the hazard probability $W(\Delta t | t)$ to measure the probability that a following extreme is about to occur in $\Delta t$ time conditioned on the last extreme happened at $t$ time in the past. \cite{Sornette-Knopoff-1997-BSSA, Bogachev-Eichner-Bunde-2007-PRL} have given the definition of hazard probability, such that, 
\begin{equation}
 W(\Delta{t}|t)=\frac{\int_t^{t+\Delta{t}}p(\tau){\rm{d}}
   \tau}{\int_t^{\infty}p(\tau){\rm{d}} \tau},  
 \label{Eq:Wq:fTau}
\end{equation}
where $p(\tau)$ corresponds to the probability density function of recurrence intervals. In the Econophysical literatures, $p(\tau)$ is approximated by the empirical distribution of recurrence intervals between historical extremes. Such hazard probability only considers the occurring time of the extremes, the potential impact of exceeding sizes and the potential dependence between recurrence interval and exceeding size are ignored, which may deteriorate the accuracy of estimating the hazard probability.

To overcome such drawbacks, we update the hazard probability in Eq.~(\ref{Eq:Wq:fTau}) by utilizing the joint distribution $f(\tau, y)$ in Eq.~(\ref{Eq:JointDis:Tau:Ey}), such that,
\begin{equation}
W_y(\Delta t|t) = \frac{\int_{t}^{t+\Delta t}\int_{0}^{y}f(\tau,y){\rm{d}}y {\rm{d}}\tau }{\int_{t}^{\infty}\int_{0}^{y}f(\tau,y) {\rm{d}}y {\rm{d}}\tau},
\label{Eq:Wq:fTauEy}
\end{equation}
where $W_y({\Delta{t}|t})$ represents the probability that a following extreme (not exceeding the last extreme) is about to occur in $\Delta t$ time conditioned on the last extreme happened at $t$ time in the past. Alternatively, we can also write Eq.~(\ref{Eq:Wq:fTauEy}) as $W_y(\Delta t|t) = h(y, c) W(\Delta{t}|t)$, where $h(y, c)$ is the correction coefficient that incorporates the influence of the exceeding size for the latest extreme and the correlation between exceeding size and recurrence interval. 

To test the accuracy of hazard probability $W_y(\Delta t|t)$, we record the number of days of correct prediction and false alarm by setting a hazard threshold $Q_p$ and compare them with factual counterparts. To avoid excessive numbers of both prediction events, we take multiple values of $Q_{p}$ varying in the interval $[0, 1]$ to calculate the false alarm rate $A$ and correct prediction rate $D$ \citep{Bogachev-Bunde-2011-PA, Jiang-Canabarro-Podobnik-Stanley-Zhou-2016-QF}, 
\begin{equation}
A=\frac{n_{10}}{n_{00}+n_{10}},~~~~D=\frac{n_{11}}{n_{01}+n_{11}}
\label{Eq:AD}
\end{equation}
where $n_{11}$ is the number of days of the extreme returns with alarm,  $n_{00}$ is the number of days without extreme yield yet alarm,  $n_{01}$ is the number of days missing the extreme yield, and $n_{10}$ is the number of days with false alarm. According to the different threshold $Q_{p}$, we can draw ROC curve with $A$ for horizontal axis and $D$ for the longitudinal axis to judge the performance of warning model \citep{Bogachev-Bunde-2009-EPL, Bogachev-Bunde-2009-PRE, Bogachev-Bunde-2011-PA, Bogachev-Kireenkov-Nifontov-Bunde-2009-NJP, Jiang-Canabarro-Podobnik-Stanley-Zhou-2016-QF, Jiang-Wang-Canabarro-Podobnik-Xie-Stanley-Zhou-2018-QF}. ROC curve above the diagonal illustrates a better performance than the benchmark of random guess. Usually, the area under the ROC curve measures the predicting performance. Following \cite{Jiang-Canabarro-Podobnik-Stanley-Zhou-2016-QF}, we also use the area under the ROC curve in the range of $0 \le A \le 0.3$ to assess the power of our model,
\begin{equation}
{\rm{AUC}}_m = \int_0^{0.3} D(A) {\rm{d}} A, 
\label{Eq:AUCm}
\end{equation}
where ${\rm{AUC}}_m$ allows us to put more focus on predicting results with low false alarm.

For a given series, we regard the last 30\% of the data for out-of-sample tests and the other 70\% of the data are used for in-sample training. The procedure for predicting extreme returns greater than $y$ is listed in the following, 
\begin{description}
\item[Step 1] The extremes are identified with a given threshold in the in-sample period, which allows us to estimate the recurrence intervals and exceeding sizes. The model parameters, including the distribution parameters of recurrence intervals and exceeding sizes, as well as the copula parameters, are estimated. 
\item[Step 2] The estimated parameters are used to calculate the hazard probability $W(\Delta t, y~|t)$ at the ending time of in-sample period. 
\item[Step 3] The in-sample period is expanded by one day and redo Step 1 and 2 till the whole data are included in the in-sample set.
\item[Step 4] The false alarm rates $A$ and correct prediction rates $D$ are estimated by comparing predicting extremes and actual extremes in in-sample and out-of-sample periods.  
\item[Step 5] By presenting the ROC curve through plotting $D$ with respect to $A$, the performance measure is estimated according to Eq.~(\ref{Eq:AUCm}).
\end{description}

\begin{figure}[htbp]
\centering
\includegraphics[width=16cm]{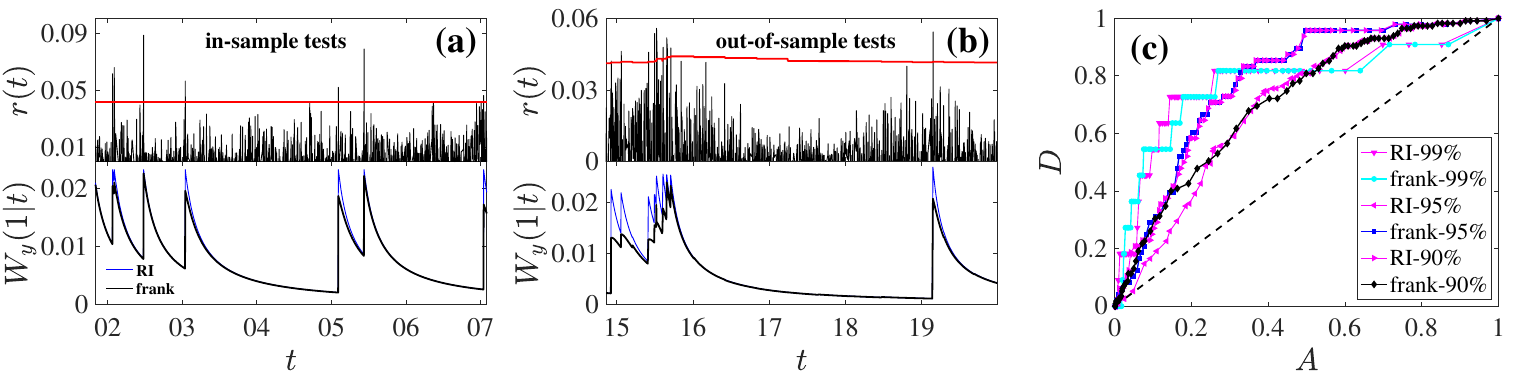}   
\caption{\label{Fig:Predict:Frank} Results of forecasting positive and negative extremes based on the Frank copula function. The positive and negative extremes are defined as the returns above the quantile of 90\%, 95\%, 99\% and the returns below the quantile of 1\%, 5\%, 10\%, respectively. The predicting parameter $\Delta t$ is set as 1 day for all analysis. (a) In-sample tests of extreme positive returns for SSEC. The extreme returns are shown in upper panel and the estimated in-sample hazard probability is plotted in lower panel. (b) Out-of-sample tests of extreme positive returns for SSEC. The extreme returns are shown in upper panel and the estimated out-of-sample hazard probability is plotted in lower panel. (c ) Out-of-sample ROC curves of predicting extreme positive returns with different quantile thresholds for SSEC.}
\end{figure}

We thus apply our model to test the performance of predicting the positive and negative extremes in SSEC, DJIA, and WTI for $\Delta t = 1$ day. The positive extremes are defined above the quantile of 90\%, 95\%, 99\% and the negative extreme are defined below the quantile of 1\%, 5\%, 10\%. Fig.~\ref{Fig:Predict:Frank} (a) plots a subseries of the positive returns (upper panel) and hazard probability $W_y(1|t)$ for in-sample tests. Frank copula is employed to estimate the hazard probability $W_y(1|t)$. The illustration of return series and the hazard probability are plotted in Fig.~\ref{Fig:Predict:Frank} (b). For comparison, the hazard probability $W(1|t)$ estimated based on the distribution of recurrence intervals (Eq.~(\ref{Eq:Wq:fTau})) is also shown in Figs.~\ref{Fig:Predict:Frank} (a) and (b). One can see that there is no obvious difference between two hazard probabilities for in-sample tests. For the out-of-sample tests, $W_y(1|t)$ outperforms $W(1|t)$, such that (1) $W_y(1|t)$ in periods with clustered extremes is much higher than that in period with sparse extremes, and (2) the extreme with large size will lead to large $W_y(1|t)$. We then apply the two hazard probabilities $W(1|t)$ and $W_y(1|t)$ to predict the extremes for different indexes and test their out-of-sample predicting performance. Fig.~\ref{Fig:Predict:Frank} (c) illustrates the ROC curves of predicting positive extremes with different thresholds for SSEC. It is observed that all the ROC curves are above the diagonal dashed line $D = A$ (random guess line), indicating that our predictions are better than the random guess. Furthermore, the ROC curves of $W_y(1|t)$ are above the ROC curves of $W(1|t)$ while predicting the extreme of the same threshold, suggesting that the predicting power of $W_y(1|t)$ is better than that of $W(1|t)$. This result means that joint model of exceeding sizes and recurrence intervals can improve the accuracy of hazard probability, thus increase the predicting performance. We also find that with the increase of quantile threshold, the predicting power of both hazard probabilities increases as well, implying that the extremes with high thresholds are easier to predict than the extremes with low threshold. 

The area under the ROC curve in the range $[0, 0.3]$ (${\rm{AUC}}_m$) is employed to quantitatively describe the performance of predicting the extremes for different thresholds and different indexes \citep{Jiang-Canabarro-Podobnik-Stanley-Zhou-2016-QF}. For each index, the predictions are conducted on six types of extremes, including the negative extremes below the quantile thresholds of 1\%, 5\%, 10\% and the positive extremes above the quantile thresholds of 90\%, 95\%, 99\%. The hazard probabilities are estimated from the distribution of recurrence intervals (Eq.~(\ref{Eq:Wq:fTau})) and the joint distribution of recurrence intervals and exceeding sizes (Eq.~(\ref{Eq:Wq:fTauEy})). In the joint distribution, we employ the Frank and AMH copula functions to connect the two marginal distributions. Table~\ref{Tab:Pred:AUC} reports the ${\rm{AUC}}_m$ for all out-of-sample predictions. We find that all the predictions are better than random guess as all the ${\rm{AUC}}_m$ are greater than 0.045. Increasing the quantile threshold for positive returns or decreasing the quantile threshold for negative returns, ${\rm{AUC}}_m$ increases, indicating this predicting framework is more suitable for forecasting the extreme tail events. It is also observed that the two hazard probabilities estimated from Eq.~(\ref{Eq:Wq:fTauEy}) give comparable predicting accuracy and generally present better prediction than that estimated from Eq.~(\ref{Eq:Wq:fTau}), suggesting that introducing exceeding size and correlation between exceeding size and recurrence intervals into recurrence interval analysis can effectively improve the predicting performance. For the same hazard probability and same index, the difference in predicting the six types of extremes can be attributed to the fact that the corresponding recurrence intervals and exceeding size have different memory behaviors. Therefore, the predicting model still can be improved by taking such memory behaviors into account.

\begin{table}[htbp]
\setlength\tabcolsep{2.6pt}
\footnotesize
\centering
 \caption{\label{Tab:Pred:AUC} Results of ${\rm{AUC}}_m$ for out-of-sample tests. The negative extreme returns (respectively, positive extreme returns) are determined according to the quantile thresholds of 1\%, 5\%, and 10\% (90\%, 95\%, and 99\%). The hazard probability used for prediction is estimated from the distribution of recurrence intervals (Eq.~(\ref{Eq:Wq:fTau})) and the joint distribution of recurrence intervals and exceeding sizes (Eq.~(\ref{Eq:Wq:fTauEy})). The Frank and AHM copulas are used to connect the two marginal distributions of recurrence intervals and exceeding sizes. }
 \medskip
  \begin{tabular}{cccccccccccccccccccccc}
\toprule
& & \multicolumn{6}{c}{SSEC} & & \multicolumn{ 6}{c}{DJIA} & & \multicolumn{ 6}{c}{WTI} \\
\cline{3-8}\cline{10-15}\cline{17-22}
& & 1\% & 5\% & 10\% & 90\% & 95\% & 99\% && 1\% & 5\% & 10\% & 90\% & 95\% & 99\% && 1\% & 5\% & 10\% & 90\% & 95\% & 99\% \\
\midrule
RI    & & 0.157 & 0.141 & 0.103 & 0.085 & 0.126 & 0.172 && 0.178 & 0.095 & 0.082 & 0.059 & 0.082 & 0.171 && 0.094 & 0.082 & 0.066 & 0.087 & 0.126 & 0.110 \\
Frank & & 0.158 & 0.150 & 0.124 & 0.104 & 0.125 & 0.170 && 0.180 & 0.105 & 0.091 & 0.067 & 0.100 & 0.172 && 0.092 & 0.101 & 0.073 & 0.101 & 0.126 & 0.111 \\
AMH   & & 0.157 & 0.150 & 0.123 & 0.101 & 0.125 & 0.172 && 0.180 & 0.104 & 0.091 & 0.068 & 0.095 & 0.174 && 0.092 & 0.098 & 0.073 & 0.102 & 0.130 & 0.111 \\
\bottomrule
\end{tabular} 
\end{table}

\section{Conclusion}

Based on three daily data sets, including the Shanghai Stock Exchange Composite (SSEC) index, the WTI Crude Oil Prices (WTI) price, and the Dow Jones Industrial Average (DJIA) index, we have developed a new framework of predicting financial extremes by utilizing the joint of distribution of the recurrence intervals and the exceeding sizes. We find that the recurrence intervals follow a $q$-exponential distribution and the exceeding sizes are well fitted by the generalized Pareto distribution. Furthermore, these two variables exhibit a significantly negative correlation for the three indexes, which motivate us to introduce the copula function to model the joint distribution of recurrence intervals and exceeding sizes. By extending the hazard function $W(1|t)$ only based on the distribution of recurrence intervals, we also incorporate the distributions of exceeding sizes and the negative correlation between recurrence intervals and exceeding sizes to the formula of hazard probability $W_y(1|t)$, which therefore improves the estimating accuracy. 

By estimating the distribution parameters and the copula parameters, we are able to calculate the hazard probabilities and perform the in-sample and out-of-sample tests on the prediction of extremes for each index. We find that there are no significant differences in extreme prediction between $W(1|t)$ and $W_y(1|t)$ for in-sample tests. For out-of-sample tests, the predicting ability in $W_y(1|t)$ is stronger than that in $W(1|t)$, evidenced by that the ROC curves of $W_y(1|t)$ are above the ROC curves of $W(1|t)$ while predicting the same type of extreme. This means that our new model has better predicting performance. Similar to the prediction based on the distribution of recurrence intervals \citep{Jiang-Wang-Canabarro-Podobnik-Xie-Stanley-Zhou-2018-QF}, our new model also favorites to predict the extremes with higher threshold for positive returns or lower threshold for negative returns. Our results not only shed a new light on understanding the occurrence of extremes in financial markets, but also provide an better framework to predict financial extremes. 

\section*{Acknowledgements}

This work was partly supported by the National Natural Science Foundation of China (U1811462, 71532009, 91746108, 71871088), the Shanghai Philosophy and Social Science Fund Project (2017BJB006), the Program of Shanghai Young Top-notch Talent (2018), the Shanghai Outstanding Academic Leaders Plan, and the Fundamental Research Funds for the Central Universities.

\bibliographystyle{elsarticle-harv}
\bibliography{E:/Papers/Auxiliary/Bibliography}

\begin{thebibliography}{59}
\expandafter\ifx\csname natexlab\endcsname\relax\def\natexlab#1{#1}\fi
\providecommand{\url}[1]{\texttt{#1}}
\providecommand{\href}[2]{#2}
\providecommand{\path}[1]{#1}
\providecommand{\DOIprefix}{doi:}
\providecommand{\ArXivprefix}{arXiv:}
\providecommand{\URLprefix}{URL: }
\providecommand{\Pubmedprefix}{pmid:}
\providecommand{\doi}[1]{\href{http://dx.doi.org/#1}{\path{#1}}}
\providecommand{\Pubmed}[1]{\href{pmid:#1}{\path{#1}}}
\providecommand{\bibinfo}[2]{#2}
\ifx\xfnm\relax \def\xfnm[#1]{\unskip,\space#1}\fi
\bibitem[{Acharya et~al.(2017)Acharya, Pedersen, Philippon and
  Richardson}]{Acharya-Pedersen-Philippon-Richardson-2017-RFS}
\bibinfo{author}{Acharya, V.V.}, \bibinfo{author}{Pedersen, L.H.},
  \bibinfo{author}{Philippon, T.}, \bibinfo{author}{Richardson, M.},
  \bibinfo{year}{2017}.
\newblock \bibinfo{title}{Measuring systemic risk}.
\newblock \bibinfo{journal}{Rev. Financ. Stud.} \bibinfo{volume}{30},
  \bibinfo{pages}{2--47}.
\newblock \DOIprefix\doi{10.1093/rfs/hhw088}.
\bibitem[{Adrian and Brunnermeier(2016)}]{Adrian-Brunnermeier-2011-AER}
\bibinfo{author}{Adrian, T.}, \bibinfo{author}{Brunnermeier, M.K.},
  \bibinfo{year}{2016}.
\newblock \bibinfo{title}{{CoVaR}}.
\newblock \bibinfo{journal}{Amer. Econ. Rev.} \bibinfo{volume}{106}.
\newblock \DOIprefix\doi{10.1257/aer.20120555}.
\bibitem[{Banulescu and Dumitrescu(2015)}]{Banulescu-Dumitrescu-2015-JBF}
\bibinfo{author}{Banulescu, G.D.}, \bibinfo{author}{Dumitrescu, E.I.},
  \bibinfo{year}{2015}.
\newblock \bibinfo{title}{{Which are the SIFIs? A Component Expected Shortfall
  approach to systemic risk}}.
\newblock \bibinfo{journal}{J. Bank. Financ.} \bibinfo{volume}{50},
  \bibinfo{pages}{575--588}.
\newblock \DOIprefix\doi{10.1016/j.jbankfin.2014.01.037}.
\bibitem[{Bogachev and Bunde(2009a)}]{Bogachev-Bunde-2009-PRE}
\bibinfo{author}{Bogachev, M.I.}, \bibinfo{author}{Bunde, A.},
  \bibinfo{year}{2009}a.
\newblock \bibinfo{title}{{Improved risk estimation in multifractal records:
  Application to the value at risk in finance}}.
\newblock \bibinfo{journal}{Phys. Rev. E} \bibinfo{volume}{80},
  \bibinfo{pages}{026131}.
\newblock \DOIprefix\doi{10.1103/PhysRevE.80.026131}.
\bibitem[{Bogachev and Bunde(2009b)}]{Bogachev-Bunde-2009-EPL}
\bibinfo{author}{Bogachev, M.I.}, \bibinfo{author}{Bunde, A.},
  \bibinfo{year}{2009}b.
\newblock \bibinfo{title}{{On the occurrence and predictability of overloads in
  telecommunication networks}}.
\newblock \bibinfo{journal}{EPL (Europhys. Lett.)} \bibinfo{volume}{86},
  \bibinfo{pages}{66002}.
\newblock \DOIprefix\doi{10.1209/0295-5075/86/66002}.
\bibitem[{Bogachev and Bunde(2011)}]{Bogachev-Bunde-2011-PA}
\bibinfo{author}{Bogachev, M.I.}, \bibinfo{author}{Bunde, A.},
  \bibinfo{year}{2011}.
\newblock \bibinfo{title}{On the predictability of extreme events in records
  with linear and nonlinear long-range memory: Efficiency and noise
  robustness}.
\newblock \bibinfo{journal}{Physica A} \bibinfo{volume}{390},
  \bibinfo{pages}{2240--2250}.
\newblock \DOIprefix\doi{10.1016/j.physa.2011.02.024}.
\bibitem[{Bogachev et~al.(2007)Bogachev, Eichner and
  Bunde}]{Bogachev-Eichner-Bunde-2007-PRL}
\bibinfo{author}{Bogachev, M.I.}, \bibinfo{author}{Eichner, J.F.},
  \bibinfo{author}{Bunde, A.}, \bibinfo{year}{2007}.
\newblock \bibinfo{title}{{Effect of nonlinear correlations on the statistics
  of return intervals in multifractal data sets}}.
\newblock \bibinfo{journal}{Phys. Rev. Lett.} \bibinfo{volume}{99},
  \bibinfo{pages}{240601}.
\newblock \DOIprefix\doi{10.1103/PhysRevLett.99.240601}.
\bibitem[{Bogachev et~al.(2009)Bogachev, Kireenkov, Nifontov and
  Bunde}]{Bogachev-Kireenkov-Nifontov-Bunde-2009-NJP}
\bibinfo{author}{Bogachev, M.I.}, \bibinfo{author}{Kireenkov, I.S.},
  \bibinfo{author}{Nifontov, E.M.}, \bibinfo{author}{Bunde, A.},
  \bibinfo{year}{2009}.
\newblock \bibinfo{title}{{Statistics of return intervals between long
  heartbeat intervals and their usability for online prediction of disorders}}.
\newblock \bibinfo{journal}{New J. Phys.} \bibinfo{volume}{11},
  \bibinfo{pages}{063036}.
\newblock \DOIprefix\doi{10.1088/1367-2630/11/6/063036}.
\bibitem[{Boubaker and Sghaier(2013)}]{Boubaker-Sghaier-2013-JBF}
\bibinfo{author}{Boubaker, H.}, \bibinfo{author}{Sghaier, N.},
  \bibinfo{year}{2013}.
\newblock \bibinfo{title}{Portfolio optimization in the presence of dependent
  financial returns with long memory: A copula based approach}.
\newblock \bibinfo{journal}{J. Bank. Financ.} \bibinfo{volume}{37},
  \bibinfo{pages}{361--377}.
\newblock \DOIprefix\doi{10.1016/j.jbankfin.2012.09.006}.
\bibitem[{Boubaker and Sghaier(2017)}]{Oh-Patton-2017-JBES}
\bibinfo{author}{Boubaker, H.}, \bibinfo{author}{Sghaier, N.},
  \bibinfo{year}{2017}.
\newblock \bibinfo{title}{Modeling dependence in high dimensions with factor
  copulas}.
\newblock \bibinfo{journal}{J. Bus. Econ. Stat.} \bibinfo{volume}{35},
  \bibinfo{pages}{139--154}.
\newblock \DOIprefix\doi{10.1080/07350015.2015.1062384}.
\bibitem[{Chicheportiche and
  Chakraborti(2013)}]{Chicheportiche-Chakraborti-2013-XXX}
\bibinfo{author}{Chicheportiche, R.}, \bibinfo{author}{Chakraborti, A.},
  \bibinfo{year}{2013}.
\newblock \bibinfo{title}{{A model-free characterization of recurrences in
  stationary time series}}.
\newblock \bibinfo{note}{Http://arxiv.org/abs/1302.3704}.
\bibitem[{Chicheportiche and
  Chakraborti(2014)}]{Chicheportiche-Chakraborti-2014-PRE}
\bibinfo{author}{Chicheportiche, R.}, \bibinfo{author}{Chakraborti, A.},
  \bibinfo{year}{2014}.
\newblock \bibinfo{title}{Copulas and time series with long-ranged
  dependencies}.
\newblock \bibinfo{journal}{Phys. Rev. E} \bibinfo{volume}{89},
  \bibinfo{pages}{042117}.
\newblock \DOIprefix\doi{10.1103/PhysRevE.89.042117}.
\bibitem[{Chollete et~al.(2011)Chollete, de~la Pe{\~n}a and
  Lu}]{Chollete-delaPena-Lu-2011-JBF}
\bibinfo{author}{Chollete, L.}, \bibinfo{author}{de~la Pe{\~n}a, V.},
  \bibinfo{author}{Lu, C.C.}, \bibinfo{year}{2011}.
\newblock \bibinfo{title}{{International diversification: A copula approach}}.
\newblock \bibinfo{journal}{J. Bank. Financ.} \bibinfo{volume}{35},
  \bibinfo{pages}{403--417}.
\newblock \DOIprefix\doi{10.1016/j.jbankfin.2010.08.020}.
\bibitem[{Chu(2011)}]{Chu-2011-JBF}
\bibinfo{author}{Chu, B.}, \bibinfo{year}{2011}.
\newblock \bibinfo{title}{{Recovering copulas from limited information and an
  application to asset allocation}}.
\newblock \bibinfo{journal}{J. Bank. Financ.} \bibinfo{volume}{35},
  \bibinfo{pages}{1824--1842}.
\newblock \DOIprefix\doi{10.1016/j.jbankfin.2010.12.011}.
\bibitem[{Corradi et~al.(2012)Corradi, Distaso and
  Fernandes}]{Corradi-Distaso-Fernandes-2012-JEm}
\bibinfo{author}{Corradi, V.}, \bibinfo{author}{Distaso, W.},
  \bibinfo{author}{Fernandes, M.}, \bibinfo{year}{2012}.
\newblock \bibinfo{title}{{International market links and volatility
  transmission}}.
\newblock \bibinfo{journal}{J. Econometr.} \bibinfo{volume}{170},
  \bibinfo{pages}{117--141}.
\newblock \DOIprefix\doi{10.1016/j.jeconom.2012.03.003}.
\bibitem[{Cotter and Suurlaht(2019)}]{Cotter-Suurlaht-2019-EJF}
\bibinfo{author}{Cotter, J.}, \bibinfo{author}{Suurlaht, A.},
  \bibinfo{year}{2019}.
\newblock \bibinfo{title}{{Spillovers in risk of financial institutions}}.
\newblock \bibinfo{journal}{Eur. J. Financ.} \bibinfo{volume}{25},
  \bibinfo{pages}{1765--1792}.
\newblock \DOIprefix\doi{10.1080/1351847X.2019.1635897}.
\bibitem[{Cumperayot and Kouwenberg(2013)}]{Cumperayot-Kouwenberg-2013-JIMF}
\bibinfo{author}{Cumperayot, P.}, \bibinfo{author}{Kouwenberg, R.},
  \bibinfo{year}{2013}.
\newblock \bibinfo{title}{{Early warning systems for currency crises: A
  multivariate extreme value approach}}.
\newblock \bibinfo{journal}{J. Int. Money Financ.} \bibinfo{volume}{36},
  \bibinfo{pages}{151--171}.
\newblock \DOIprefix\doi{10.1016/j.jimonfin.2013.03.008}.
\bibitem[{Da~Fonseca and Ignatieva(2018)}]{Fonseca-Ignatieva-2018-AE}
\bibinfo{author}{Da~Fonseca, J.}, \bibinfo{author}{Ignatieva, K.},
  \bibinfo{year}{2018}.
\newblock \bibinfo{title}{{Volatility spillovers and connectedness among credit
  default swap sector indexes}}.
\newblock \bibinfo{journal}{Appl. Econ.} \bibinfo{volume}{50},
  \bibinfo{pages}{3923--3936}.
\newblock \DOIprefix\doi{10.1080/00036846.2018.1430344}.
\bibitem[{Elyasiani et~al.(2015)Elyasiani, Kalotychou, Staikouras and
  Zhao}]{Elyasiani-Kalotychou-Staikouras-Zhao-2015-JFSR}
\bibinfo{author}{Elyasiani, E.}, \bibinfo{author}{Kalotychou, E.},
  \bibinfo{author}{Staikouras, S.K.}, \bibinfo{author}{Zhao, G.},
  \bibinfo{year}{2015}.
\newblock \bibinfo{title}{{Return and volatility spillover among banks and
  insurers: Evidence from pre-crisis and crisis periods}}.
\newblock \bibinfo{journal}{J. Financ. Services Res.} \bibinfo{volume}{48},
  \bibinfo{pages}{21--52}.
\newblock \DOIprefix\doi{10.1007/s10693-014-0200-z}.
\bibitem[{Engle and Ruan(2019)}]{Engle-Ruan-2019-PNAS}
\bibinfo{author}{Engle, R.F.}, \bibinfo{author}{Ruan, T.Y.},
  \bibinfo{year}{2019}.
\newblock \bibinfo{title}{{Measuring the probability of a financial crisis}}.
\newblock \bibinfo{journal}{Proc. Natl. Acad. Sci. U.S.A.}
  \bibinfo{volume}{116}, \bibinfo{pages}{18341--18346}.
\newblock \DOIprefix\doi{10.1073/pnas.1903879116}.
\bibitem[{Fry-McKibbin et~al.(2014)Fry-McKibbin, Martin and
  Tang}]{McKibbin-Martin-Tang-2014-JBF}
\bibinfo{author}{Fry-McKibbin, R.}, \bibinfo{author}{Martin, V.L.},
  \bibinfo{author}{Tang, C.}, \bibinfo{year}{2014}.
\newblock \bibinfo{title}{{Financial contagion and asset pricing}}.
\newblock \bibinfo{journal}{J. Bank. Financ.} \bibinfo{volume}{47},
  \bibinfo{pages}{296--308}.
\newblock \DOIprefix\doi{10.1016/j.jbankfin.2014.05.002}.
\bibitem[{Girardi and Erg{\"u}n(2013)}]{Girardi-Ergun-2013-JBF}
\bibinfo{author}{Girardi, G.}, \bibinfo{author}{Erg{\"u}n, A.T.},
  \bibinfo{year}{2013}.
\newblock \bibinfo{title}{Systemic risk measurement: {M}ultivariate {GARCH}
  estimation of {CoVaR}}.
\newblock \bibinfo{journal}{J. Bank. Financ.} \bibinfo{volume}{37},
  \bibinfo{pages}{3169--3180}.
\bibitem[{Greco et~al.(2008)Greco, Sorriso-Valvo, Carbone and
  Cidone}]{Greco-SorrisoValvo-Carbone-Cidone-2008-PA}
\bibinfo{author}{Greco, A.}, \bibinfo{author}{Sorriso-Valvo, L.},
  \bibinfo{author}{Carbone, V.}, \bibinfo{author}{Cidone, S.},
  \bibinfo{year}{2008}.
\newblock \bibinfo{title}{{Waiting time distributions of the volatility in the
  Italian MIB30 index: Clustering or Poisson functions?}}
\newblock \bibinfo{journal}{Physica A} \bibinfo{volume}{387},
  \bibinfo{pages}{4272--4284}.
\newblock \DOIprefix\doi{10.1016/j.physa.2008.03.007}.
\bibitem[{Gresnigt et~al.(2015)Gresnigt, Kole and
  Franses}]{Gresnigt-Kole-Franses-2015-JBF}
\bibinfo{author}{Gresnigt, F.}, \bibinfo{author}{Kole, E.},
  \bibinfo{author}{Franses, P.H.}, \bibinfo{year}{2015}.
\newblock \bibinfo{title}{Interpreting financial market crashes as earthquakes:
  A new early warning system for medium term crashes}.
\newblock \bibinfo{journal}{J. Bank. Financ.} \bibinfo{volume}{56},
  \bibinfo{pages}{123--139}.
\newblock \DOIprefix\doi{10.1016/j.jbankfin.2015.03.003}.
\bibitem[{Ji et~al.(2018)Ji, Bouri, Roubaud and
  H.}]{Ji-Bouri-Roubaud-Shahzad-2018-EE}
\bibinfo{author}{Ji, Q.}, \bibinfo{author}{Bouri, E.},
  \bibinfo{author}{Roubaud, D.}, \bibinfo{author}{H., S.S.J.},
  \bibinfo{year}{2018}.
\newblock \bibinfo{title}{{Risk spillover between energy and agricultural
  commodity markets: A dependence-switching CoVaR-copula model}}.
\newblock \bibinfo{journal}{Energy Econ.} \bibinfo{volume}{75},
  \bibinfo{pages}{14--27}.
\newblock \DOIprefix\doi{10.1016/j.eneco.2018.08.015}.
\bibitem[{Ji et~al.(2019)Ji, Liu and Fan}]{Ji-Liu-Fan-2019-EE}
\bibinfo{author}{Ji, Q.}, \bibinfo{author}{Liu, B.Y.}, \bibinfo{author}{Fan,
  Y.}, \bibinfo{year}{2019}.
\newblock \bibinfo{title}{{Risk dependence of CoVaR and structural change
  between oil prices and exchange rates: A time-varying copula model}}.
\newblock \bibinfo{journal}{Energy Econ.} \bibinfo{volume}{77},
  \bibinfo{pages}{80--92}.
\newblock \DOIprefix\doi{10.1016/j.eneco.2018.07.012}.
\bibitem[{Jiang et~al.(2016)Jiang, Canabarro, Podobnik, Stanely and
  Zhou}]{Jiang-Canabarro-Podobnik-Stanley-Zhou-2016-QF}
\bibinfo{author}{Jiang, Z.Q.}, \bibinfo{author}{Canabarro, A.A.},
  \bibinfo{author}{Podobnik, B.}, \bibinfo{author}{Stanely, H.E.},
  \bibinfo{author}{Zhou, W.X.}, \bibinfo{year}{2016}.
\newblock \bibinfo{title}{{Early warning of large volatilities based on
  recurrence interval analysis in Chinese stock markets}}.
\newblock \bibinfo{journal}{Quant. Financ.} \bibinfo{volume}{16},
  \bibinfo{pages}{1713--1724}.
\newblock \DOIprefix\doi{10.1080/14697688.2016.1175656}.
\bibitem[{Jiang et~al.(2018)Jiang, Wang, Cananarro, Podobnik, Xie, Stanley and
  Zhou}]{Jiang-Wang-Canabarro-Podobnik-Xie-Stanley-Zhou-2018-QF}
\bibinfo{author}{Jiang, Z.Q.}, \bibinfo{author}{Wang, G.J.},
  \bibinfo{author}{Cananarro, A.}, \bibinfo{author}{Podobnik, B.},
  \bibinfo{author}{Xie, C.}, \bibinfo{author}{Stanley, H.E.},
  \bibinfo{author}{Zhou, W.X.}, \bibinfo{year}{2018}.
\newblock \bibinfo{title}{{Short term prediction of extreme returns based on
  the recurrence interval analysis}}.
\newblock \bibinfo{journal}{Quant. Financ.} \bibinfo{volume}{18},
  \bibinfo{pages}{353--370}.
\newblock \DOIprefix\doi{10.1080/14697688.2017.1373843}.
\bibitem[{Jung and Maderitsch(2014)}]{Jung-Maderitsch-2014-JBF}
\bibinfo{author}{Jung, R.C.}, \bibinfo{author}{Maderitsch, R.},
  \bibinfo{year}{2014}.
\newblock \bibinfo{title}{{Structural breaks in volatility spillovers between
  international financial markets: Contagion or mere interdependence?}}
\newblock \bibinfo{journal}{J. Bank. Financ.} \bibinfo{volume}{47},
  \bibinfo{pages}{331--342}.
\newblock \DOIprefix\doi{10.1016/j.jbankfin.2013.12.023}.
\bibitem[{Kenourgios et~al.(2011)Kenourgios, Samitas and
  Paltalidis}]{Kenourgios-Samitas-Paltalidis-2011-JIFMIM}
\bibinfo{author}{Kenourgios, D.}, \bibinfo{author}{Samitas, A.},
  \bibinfo{author}{Paltalidis, N.}, \bibinfo{year}{2011}.
\newblock \bibinfo{title}{{Financial crises and stock market contagion in a
  multivariate time-varying asymmetric framework}}.
\newblock \bibinfo{journal}{J. Int. Financ. Mark. Inst. Money}
  \bibinfo{volume}{21}, \bibinfo{pages}{92--106}.
\newblock \DOIprefix\doi{10.1016/j.intfin.2010.08.005}.
\bibitem[{Lee et~al.(2006)Lee, Lee and Rikvold}]{Lee-Lee-Rikvold-2006-JKPS}
\bibinfo{author}{Lee, J.W.}, \bibinfo{author}{Lee, K.E.},
  \bibinfo{author}{Rikvold, P.A.}, \bibinfo{year}{2006}.
\newblock \bibinfo{title}{{Waiting-time distribution for Korean stock-market
  index KOSPI}}.
\newblock \bibinfo{journal}{J. Korean Phys. Soc.} \bibinfo{volume}{48},
  \bibinfo{pages}{S123--S126}.
\bibitem[{Li et~al.(2011)Li, Wang, Havlin and
  Stanley}]{Li-Wang-Havlin-Stanley-2011-PRE}
\bibinfo{author}{Li, W.}, \bibinfo{author}{Wang, F.Z.},
  \bibinfo{author}{Havlin, S.}, \bibinfo{author}{Stanley, H.E.},
  \bibinfo{year}{2011}.
\newblock \bibinfo{title}{{Financial factor influence on scaling and memory of
  trading volume in stock market}}.
\newblock \bibinfo{journal}{Phys. Rev. E} \bibinfo{volume}{84},
  \bibinfo{pages}{046112}.
\newblock \DOIprefix\doi{10.1103/PhysRevE.84.046112}.
\bibitem[{Li and Wei(2018)}]{Li-Wei-2018-EE}
\bibinfo{author}{Li, X.F.}, \bibinfo{author}{Wei, Y.}, \bibinfo{year}{2018}.
\newblock \bibinfo{title}{{The dependence and risk spillover between crude oil
  market and China stock market: New evidence from a variational mode
  decomposition-based copula method}}.
\newblock \bibinfo{journal}{Energy Econ.} \bibinfo{volume}{74},
  \bibinfo{pages}{565--581}.
\newblock \DOIprefix\doi{10.1016/j.eneco.2018.07.011}.
\bibitem[{Liu et~al.(2017)Liu, Ji and Fan}]{Liu-Ji-Fan-2017-EE}
\bibinfo{author}{Liu, B.Y.}, \bibinfo{author}{Ji, Q.}, \bibinfo{author}{Fan,
  Y.}, \bibinfo{year}{2017}.
\newblock \bibinfo{title}{{Dynamic return-volatility dependence and risk
  measure of CoVaR in the oil market: A time-varying mixed copula model}}.
\newblock \bibinfo{journal}{Energy Econ.} \bibinfo{volume}{68},
  \bibinfo{pages}{53--65}.
\newblock \DOIprefix\doi{10.1016/j.eneco.2017.09.011}.
\bibitem[{Ludescher and Bunde(2014)}]{Ludescher-Bunde-2014-PRE}
\bibinfo{author}{Ludescher, J.}, \bibinfo{author}{Bunde, A.},
  \bibinfo{year}{2014}.
\newblock \bibinfo{title}{Universal behavior of the interoccurrence times
  between losses in financial markets: Independence of the time resolution}.
\newblock \bibinfo{journal}{Phys. Rev. E} \bibinfo{volume}{90},
  \bibinfo{pages}{062809}.
\newblock \DOIprefix\doi{10.1103/PhysRevE.90.062809}.
\bibitem[{Ludescher et~al.(2011)Ludescher, Tsallis and
  Bunde}]{Ludescher-Tsallis-Bunde-2011-EPL}
\bibinfo{author}{Ludescher, J.}, \bibinfo{author}{Tsallis, C.},
  \bibinfo{author}{Bunde, A.}, \bibinfo{year}{2011}.
\newblock \bibinfo{title}{{Universal behaviour of interoccurrence times between
  losses in financial markets: An analytical description}}.
\newblock \bibinfo{journal}{EPL (Europhys. Lett.)} \bibinfo{volume}{95},
  \bibinfo{pages}{68002}.
\newblock \DOIprefix\doi{10.1209/0295-5075/95/68002}.
\bibitem[{Malevergne et~al.(2006)Malevergne, Pisarenko and
  Sornette}]{Malevergne-Pisarenko-Sornette-2006-AFE}
\bibinfo{author}{Malevergne, Y.}, \bibinfo{author}{Pisarenko, V.},
  \bibinfo{author}{Sornette, D.}, \bibinfo{year}{2006}.
\newblock \bibinfo{title}{{On the power of generalized extreme value (GEV) and
  generalized Pareto distribution (GPD) estimators for empirical distributions
  of stock returns}}.
\newblock \bibinfo{journal}{Appl. Financial Econ.} \bibinfo{volume}{16},
  \bibinfo{pages}{271--289}.
\newblock \DOIprefix\doi{10.1080/09603100500391008}.
\bibitem[{Mensi et~al.(2017)Mensi, Hammoudeh, Shahzad and
  Shahbaz}]{Mensi-Hammoudeh-Shahzad-Shahbaz-2017-JBF}
\bibinfo{author}{Mensi, W.}, \bibinfo{author}{Hammoudeh, S.},
  \bibinfo{author}{Shahzad, S.J.H.}, \bibinfo{author}{Shahbaz, M.},
  \bibinfo{year}{2017}.
\newblock \bibinfo{title}{Modeling systemic risk and dependence structure
  between oil and stock markets using a variational mode decomposition-based
  copula method}.
\newblock \bibinfo{journal}{J. Bank. Financ.} \bibinfo{volume}{75},
  \bibinfo{pages}{258--279}.
\bibitem[{Nikoloulopoulos et~al.(2012)Nikoloulopoulos, Joe and
  Li}]{Nikoloulopoulos-Joe-Li-2012-CSDA}
\bibinfo{author}{Nikoloulopoulos, A.K.}, \bibinfo{author}{Joe, H.},
  \bibinfo{author}{Li, H.J.}, \bibinfo{year}{2012}.
\newblock \bibinfo{title}{Vine copulas with asymmetric tail dependence and
  applications to financial return data}.
\newblock \bibinfo{journal}{Comput. Stat. Data Anal.} \bibinfo{volume}{56},
  \bibinfo{pages}{3659--3673}.
\newblock \DOIprefix\doi{10.1016/j.csda.2010.07.016}.
\bibitem[{Ning(2010)}]{Ning-2010-JIMF}
\bibinfo{author}{Ning, C.}, \bibinfo{year}{2010}.
\newblock \bibinfo{title}{{Dependence structure between the equity market and
  the foreign exchange market – A copula approach}}.
\newblock \bibinfo{journal}{J. Int. Money Financ.} \bibinfo{volume}{29},
  \bibinfo{pages}{743--759}.
\newblock \DOIprefix\doi{10.1016/j.jimonfin.2009.12.002}.
\bibitem[{Patton(2006)}]{Patton-2006-IER}
\bibinfo{author}{Patton, A.J.}, \bibinfo{year}{2006}.
\newblock \bibinfo{title}{{Modelling asymmetric exchange rate dependence}}.
\newblock \bibinfo{journal}{Int. Econ. Rev.} \bibinfo{volume}{47},
  \bibinfo{pages}{527--556}.
\newblock \DOIprefix\doi{10.1111/j.1468-2354.2006.00387.x}.
\bibitem[{Reboredo et~al.(2014)Reboredo, Rivera-Castro and Machado~de
  Assis}]{Reboredo-Rivera-Castro-Machado-2014-QF}
\bibinfo{author}{Reboredo, J.C.}, \bibinfo{author}{Rivera-Castro, M.A.},
  \bibinfo{author}{Machado~de Assis, E.}, \bibinfo{year}{2014}.
\newblock \bibinfo{title}{{Power-law behaviour in time durations between
  extreme returns}}.
\newblock \bibinfo{journal}{Quant. Financ.} \bibinfo{volume}{14},
  \bibinfo{pages}{2171--2183}.
\newblock \DOIprefix\doi{10.1080/14697688.2013.822538}.
\bibitem[{Reboredo and Ugolini(2015)}]{Reboredo-Ugolini-2015-JIMF}
\bibinfo{author}{Reboredo, J.C.}, \bibinfo{author}{Ugolini, A.},
  \bibinfo{year}{2015}.
\newblock \bibinfo{title}{Systemic risk in {E}uropean sovereign debt markets:
  {A} {CoVaR}-copula approach}.
\newblock \bibinfo{journal}{J. Int. Money Financ.} \bibinfo{volume}{51},
  \bibinfo{pages}{214--244}.
\bibitem[{Ren and Zhou(2010a)}]{Ren-Zhou-2010-NJP}
\bibinfo{author}{Ren, F.}, \bibinfo{author}{Zhou, W.X.}, \bibinfo{year}{2010}a.
\newblock \bibinfo{title}{{Recurrence interval analysis of high-frequency
  financial returns and its application to risk estimation}}.
\newblock \bibinfo{journal}{New J. Phys.} \bibinfo{volume}{12},
  \bibinfo{pages}{075030}.
\newblock \DOIprefix\doi{10.1088/1367-2630/12/7/075030}.
\bibitem[{Ren and Zhou(2010b)}]{Ren-Zhou-2010-PRE}
\bibinfo{author}{Ren, F.}, \bibinfo{author}{Zhou, W.X.}, \bibinfo{year}{2010}b.
\newblock \bibinfo{title}{{Recurrence interval analysis of trading volumes}}.
\newblock \bibinfo{journal}{Phys. Rev. E} \bibinfo{volume}{81},
  \bibinfo{pages}{066107}.
\newblock \DOIprefix\doi{10.1103/PhysRevE.81.066107}.
\bibitem[{Santhanam and Kantz(2008)}]{Santhanam-Kantz-2008-PRE}
\bibinfo{author}{Santhanam, M.S.}, \bibinfo{author}{Kantz, H.},
  \bibinfo{year}{2008}.
\newblock \bibinfo{title}{{Return interval distribution of extreme events and
  long-term memory}}.
\newblock \bibinfo{journal}{Phys. Rev. E} \bibinfo{volume}{78},
  \bibinfo{pages}{051113}.
\newblock \DOIprefix\doi{10.1103/PhysRevE.78.051113}.
\bibitem[{Shahzad et~al.(2019)Shahzad, Bouri, Arreola-Hernandez, Roubaud and
  Bekiros}]{Shahzad-Bouri-Arreola-Roubaud-Bekiros-2019-AE}
\bibinfo{author}{Shahzad, S.J.H.}, \bibinfo{author}{Bouri, E.},
  \bibinfo{author}{Arreola-Hernandez, J.}, \bibinfo{author}{Roubaud, D.},
  \bibinfo{author}{Bekiros, S.}, \bibinfo{year}{2019}.
\newblock \bibinfo{title}{{Spillover across Eurozone credit market sectors and
  determinants}}.
\newblock \bibinfo{journal}{Appl. Econ.} \bibinfo{volume}{51},
  \bibinfo{pages}{6333--6349}.
\newblock \DOIprefix\doi{10.1080/00036846.2019.1619014}.
\bibitem[{Siburg et~al.(2015)Siburg, Stoimenov and
  Wei{\ss}}]{Siburg-Stoimenov-Weiss-2015-JBF}
\bibinfo{author}{Siburg, K.F.}, \bibinfo{author}{Stoimenov, P.},
  \bibinfo{author}{Wei{\ss}, G.N.F.}, \bibinfo{year}{2015}.
\newblock \bibinfo{title}{{Forecasting portfolio-Value-at-Risk with
  nonparametric lower tail dependence estimates}}.
\newblock \bibinfo{journal}{J. Bank. Financ.} \bibinfo{volume}{54},
  \bibinfo{pages}{129--140}.
\newblock \DOIprefix\doi{10.1016/j.jbankfin.2015.01.012}.
\bibitem[{Sklar(1959)}]{Sklar-1959-PISUP}
\bibinfo{author}{Sklar, A.}, \bibinfo{year}{1959}.
\newblock \bibinfo{title}{Fonctions de ri{\'e}partition {\'a} n dimensions et
  leurs marges}.
\newblock \bibinfo{journal}{Publications de l’Institutde Statistique de
  l’Universit{\'e} de Paris} \bibinfo{volume}{8}, \bibinfo{pages}{229--231}.
\bibitem[{Sornette and Knopoff(1997)}]{Sornette-Knopoff-1997-BSSA}
\bibinfo{author}{Sornette, D.}, \bibinfo{author}{Knopoff, L.},
  \bibinfo{year}{1997}.
\newblock \bibinfo{title}{{The paradox of the expected time until the next
  earthquake}}.
\newblock \bibinfo{journal}{Bull. Seism. Soc. Am.} \bibinfo{volume}{87},
  \bibinfo{pages}{789--798}.
\bibitem[{Suo et~al.(2015)Suo, Wang and Li}]{Suo-Wang-Li-2015-EM}
\bibinfo{author}{Suo, Y.Y.}, \bibinfo{author}{Wang, D.H.}, \bibinfo{author}{Li,
  S.P.}, \bibinfo{year}{2015}.
\newblock \bibinfo{title}{Risk estimation of csi 300 index spot and futures in
  china from a new perspective}.
\newblock \bibinfo{journal}{Econ. Model.} \bibinfo{volume}{49},
  \bibinfo{pages}{344--353}.
\newblock \DOIprefix\doi{10.1016/j.econmod.2015.05.011}.
\bibitem[{Wang and Wang(2012)}]{Wang-Wang-2012-CIE}
\bibinfo{author}{Wang, F.}, \bibinfo{author}{Wang, J.}, \bibinfo{year}{2012}.
\newblock \bibinfo{title}{{Statistical analysis and forecasting of return
  interval for SSE and model by lattice percolation system and neural
  network}}.
\newblock \bibinfo{journal}{Comput. Ind. Eng.} \bibinfo{volume}{62},
  \bibinfo{pages}{198--205}.
\newblock \DOIprefix\doi{10.1016/j.cie.2011.09.007}.
\bibitem[{Wang et~al.(2018)Wang, Xie, Zhao and
  Jiang}]{Wang-Xie-Zhao-Jiang-2018-JIFMIM}
\bibinfo{author}{Wang, G.J.}, \bibinfo{author}{Xie, C.}, \bibinfo{author}{Zhao,
  L.}, \bibinfo{author}{Jiang, Z.Q.}, \bibinfo{year}{2018}.
\newblock \bibinfo{title}{{Volatility connectedness in the Chinese banking
  system: Do state-owned commercial banks contribute more?}}
\newblock \bibinfo{journal}{J. Int. Financ. Mark. Inst. Money}
  \bibinfo{volume}{57}, \bibinfo{pages}{205--230}.
\newblock \DOIprefix\doi{10.1016/j.intfin.2018.07.008}.
\bibitem[{Wang and Wu(2019)}]{Wang-Wu-2019-EE}
\bibinfo{author}{Wang, X.X.}, \bibinfo{author}{Wu, C.F.}, \bibinfo{year}{2019}.
\newblock \bibinfo{title}{{Volatility spillovers between crude oil and Chinese
  sectoral equitymarkets: Evidence from a frequency dynamics perspective}}.
\newblock \bibinfo{journal}{Energy Econ.} \bibinfo{volume}{80},
  \bibinfo{pages}{995–1009}.
\newblock \DOIprefix\doi{10.1016/j.eneco.2019.02.019}.
\bibitem[{Wang et~al.(2013)Wang, Wu and Lai}]{Wang-Wu-Lai-2013-JBF}
\bibinfo{author}{Wang, Y.C.}, \bibinfo{author}{Wu, J.H.}, \bibinfo{author}{Lai,
  Y.H.}, \bibinfo{year}{2013}.
\newblock \bibinfo{title}{A revisit to the dependence structure between the
  stock and foreign exchange markets: A dependence-switching copula approach}.
\newblock \bibinfo{journal}{J. Bank. Financ.} \bibinfo{volume}{37},
  \bibinfo{pages}{1706--1719}.
\newblock \DOIprefix\doi{10.1016/j.jbankfin.2013.01.001}.
\bibitem[{Wu et~al.(2019)Wu, Zhang and Zhang}]{Wu-Zhang-Zhang-2019-ES}
\bibinfo{author}{Wu, F.}, \bibinfo{author}{Zhang, D.}, \bibinfo{author}{Zhang,
  Z.}, \bibinfo{year}{2019}.
\newblock \bibinfo{title}{{Connectedness and risk spillovers in China’s stock
  market: A sectoral analysis}}.
\newblock \bibinfo{journal}{Energy Source.} \bibinfo{volume}{43},
  \bibinfo{pages}{100718}.
\newblock \DOIprefix\doi{10.1016/j.ecosys.2019.100718}.
\bibitem[{Xie et~al.(2014)Xie, Jiang and Zhou}]{Xie-Jiang-Zhou-2014-EM}
\bibinfo{author}{Xie, W.J.}, \bibinfo{author}{Jiang, Z.Q.},
  \bibinfo{author}{Zhou, W.X.}, \bibinfo{year}{2014}.
\newblock \bibinfo{title}{{Extreme value statistics and recurrence intervals of
  NYMEX energy futures volatility}}.
\newblock \bibinfo{journal}{Econ. Model.} \bibinfo{volume}{36},
  \bibinfo{pages}{8--17}.
\newblock \DOIprefix\doi{10.1016/j.econmod.2013.09.011}.
\bibitem[{Yamasaki et~al.(2005)Yamasaki, Muchnik, Havlin, Bunde and
  Stanley}]{Yamasaki-Muchnik-Havlin-Bunde-Stanley-2005-PNAS}
\bibinfo{author}{Yamasaki, K.}, \bibinfo{author}{Muchnik, L.},
  \bibinfo{author}{Havlin, S.}, \bibinfo{author}{Bunde, A.},
  \bibinfo{author}{Stanley, H.E.}, \bibinfo{year}{2005}.
\newblock \bibinfo{title}{{Scaling and memory in volatility return intervals in
  financial markets}}.
\newblock \bibinfo{journal}{Proc. Natl. Acad. Sci. U.S.A.}
  \bibinfo{volume}{102}, \bibinfo{pages}{9424--9428}.
\newblock \DOIprefix\doi{10.1073/pnas.0502613102}.
\bibitem[{Zhang(2005)}]{Zhang-2005-PhDthesis}
\bibinfo{author}{Zhang, L.}, \bibinfo{year}{2005}.
\newblock \bibinfo{title}{Multivariate hydrological frequency analysis and risk
  mapping}.
\newblock Ph.D. thesis. Louisiana State University.
\newblock \URLprefix
  \url{https://digitalcommons.lsu.edu/gradschool{\_}dissertations/1351}.

\end{thebibliography}

\end{document}